\documentclass[preprint]{elsarticle}
\usepackage{graphicx}
\usepackage{txfonts}
\usepackage{color}
\usepackage{natbib}
\usepackage{tabularx}
\usepackage{lineno}
\bibpunct{(}{)}{;}{a}{}{,} % to follow the A&A style

\begin{document}

\begin{frontmatter}

\title{Estimating precipitation on early Mars using a radiative-convective
model of the atmosphere and comparison with inferred runoff from geomorphology
}

\author[dlr,lab,cnrs]{P. von Paris\corref{cor}}
\author[dlr]{A. Petau}
\author[dlr]{J.L. Grenfell}
\author[dlr]{E. Hauber}
\author[dlr]{D. Breuer}
\author[dlr]{R. Jaumann}
\author[dlr,zaa]{H. Rauer}
\author[dlr]{D. Tirsch}

\cortext[cor]{Corresponding author: email vonparis@obs.u-bordeaux1.fr, tel. +33 (0)557 77 6131}

\address[dlr]{Institut f\"{u}r Planetenforschung, Deutsches Zentrum
f\"{u}r Luft- und Raumfahrt (DLR), Rutherfordstr. 2, 12489 Berlin, Germany}
\address[lab]{now at: Univ. Bordeaux, LAB, UMR 5804, F-33270, Floirac, France}
\address[cnrs]{now at: CNRS, LAB, UMR 5804, F-33270, Floirac, France }
\address[zaa]{Zentrum f\"{u}r Astronomie und Astrophysik (ZAA), Technische
Universit\"{a}t Berlin, Hardenbergstr. 36, 10623 Berlin, Germany}

\begin{abstract}

We compare estimates of atmospheric precipitation during the Martian Noachian-Hesperian boundary 3.8 Gyr ago as calculated in a radiative-convective column model of the atmosphere with runoff values estimated from a geomorphological analysis of dendritic valley network discharge rates. In the atmospheric model, we assume CO$_2$-H$_2$O-N$_2$ atmospheres with surface pressures varying from 20\,mb to 3\,bar with input solar luminosity reduced to 75\% the modern value.

Results from the valley network analysis are of the order of a few mm d$^{-1}$ liquid water precipitation (1.5-10.6\,mm d$^{-1}$, with a median of 3.1\,mm d$^{-1}$). Atmospheric model results are much lower, from about 0.001-1\,mm d$^{-1}$ of snowfall (depending on CO$_2$ partial pressure). Hence, the atmospheric model predicts a significantly lower amount of precipitated water than estimated from the geomorphological analysis. Furthermore, global mean surface temperatures are below freezing, i.e. runoff is most likely not directly linked to precipitation. Therefore, our results strongly favor a cold early Mars with episodic snowmelt as a source for runoff. 

Our approach is challenged by mostly unconstrained parameters, e.g. greenhouse gas abundance, global meteorology (for example, clouds) and planetary parameters such as obliquity- which affect the atmospheric result - as as well as by inherent problems in estimating discharge and runoff on ancient Mars, such as a lack of knowledge on infiltration and evaporation rates and on flooding timescales, which affect the geomorphological data. Nevertheless, our work represents a first  step in combining and interpreting quantitative tools applied in early Mars atmospheric and geomorphological studies.

\end{abstract}

\end{frontmatter}

\section{Introduction}

Habitability defined as the conditions suitable for life (e.g., \citealp{mepag2005}) has become a central concept in both Solar System and exoplanet science. Early Mars is arguably the key environment to study whether habitable conditions could arise away from the Earth.

In this work we apply an atmospheric model to estimate {global mean} precipitation rates on early Mars. These are then compared with runoff rates as derived from a geomorphological data analysis of a sample of valley networks. Our main aim is not to investigate the formation of individual networks. Rather, we aim to assess (i) the probable strength of the overall hydrological cycle on early Mars in terms of the amount of precipitated water needed to form the networks, and (ii) whether the atmospheric conditions would have allowed for such a hydrological cycle, again in terms of amount of precipitated water, but also in terms of temperature (snow vs. rainfall).
 
We begin (Section \ref{atmo_back}) by discussing processes affecting atmospheric formation and composition – since these are critical for the early Mars climate hence habitability. Then we give an overview of the geomorphological valley features (Section \ref{geo_back}) observed on Mars which provide key evidence that early Mars was wet.
Section \ref{tools} presents the tools used and their constraints. Section \ref{results} presents results, comparing precipitation rates from the atmospheric model with those from the geomorphological approach. Section \ref{disc} presents a discussion and Section \ref{concl} shows conclusions.

\subsection{Background on early Mars Atmosphere }
\label{atmo_back}

Constraints on atmospheric composition and mass for the early Martian atmosphere can be obtained from a combination of outgassing and escape modeling as well as measurements of e.g. isotopic ratios of nitrogen, oxygen and carbon. Degassing during the early magma ocean phase could have led to an atmosphere of tens of bars or more, but this was probably efficiently removed very fast either during the magma-ocean phase or at the latest during the first few hundred million years due to strong solar activity (e.g., \citealp{tian2009_mars}, \citealp{lammer2013}). Later input by outgassing is likely  insufficient to form dense CO$_2$ atmospheres of the order of a few bars. Model studies suggest a maximum of about 0.5-1.5\,bar before 3.8 Gyrs (e.g., \citealp{phillips2001}, \citealp{grott2011}), with the lower value being more realistic considering the low oxygen fugacity of the Martian mantle and that crustal recycling was inefficient in Mars (e.g., \citealp{stanley2011}). However, impacts during the late heavy bombardment may have provided additional atmospheric mass (up to a few bars, e.g., \citealp{deniem2012}). Isotopic ratios (e.g., \citealp{jak2001}, \citealp{fox2010}, \citealp{gillmann2011}) and an in-situ analysis of estimated rock trajectories during explosive volcanic eruptions (e.g., \citealp{manga2012}) also suggest a denser atmosphere than today. Upper limits on early Mars atmospheric pressure of about 1\,bar have been reported recently based on crater analysis \citep{kite2014}.  A key challenge is how to remove considerable amounts of atmosphere in order to arrive at the present, thin atmosphere, since loss processes are not thought to be efficient after the Noachian period (e.g., \citealp{lammer2013}).

The early Mars atmosphere is thought to be composed mainly of CO$_2$, as suggested by outgassing models (e.g., \citealp{phillips2001}), although such studies also predict significant H$_2$O outgassing (e.g., \citealp{grott2011}). Trace gases could have been present in the atmosphere, e.g. SO$_2$ due to volcanic outgassing (e.g., \citealp{farq2000}, \citealp{halevy2007}) or O$_3$ due to atmospheric photochemistry (e.g., \citealp{selsis2002}). Atmospheric N$_2$ may have been present since its original inventory is relatively large (e.g., \citealp{mckay1989mars}). Other radiatively active gases such as CH$_4$ have also been suggested (e.g., \citealp{postawko1986}). Recent studies investigated the possibility of H$_2$-induced warming (e.g., \citealp{ramirez2014}) because H$_2$ could have been a major atmospheric constituent due to enhanced outgassing from the reduced early Mars mantle. However, most atmospheric model studies only investigated CO$_2$-H$_2$O scenarios, some with the addition of either SO$_2$, H$_2$ or N$_2$, but currently no model has used a combination of all of these gases.

Early 1D CO$_2$-H$_2$O atmospheric model studies by, e.g., \citet{kasting1991} suggested mean surface temperatures far below freezing, indicating that sustained rainfall might not be the reason for producing observed fluvial features. One possible mechanism suggested for warming early Mars includes the formation of CO$_2$ clouds (e.g., \citealp{pierre1998}, \citealp{forget1997}). However, 1D and 3D modeling studies suggested that the cloud cover would have to be nearly 100\% (e.g., \citealp{mischna2000}), which is unrealistic as found by more detailed, time-dependent 1D or 3D simulations (e.g., \citealp{Cola2003}, \citealp{wordsworth2013}). Recent radiative transfer modeling studies \citep{kitzmann2013} suggested that the overall warming effect might have been strongly overestimated.

Most recent 1D atmospheric modeling studies continue to calculate mean surface temperatures below freezing even when including the presence of additional greenhouse gases such as SO$_2$ (e.g., \citealp{tian2010}) or N$_2$ (e.g., \citealp{vparis2013marsn2}). In contrast, the new study by \citet{ramirez2014} found mean surface temperatures well above freezing upon simulating dense CO$_2$-H$_2$ atmospheres. \citet{kahre2013} speculate that a highly active dust cycle on early Mars could have warmed the surface. Dust could have warmed the surface by up to 10\,K depending on dust opacity \citep{forget2013}. A reduction in surface albedo (e.g. due to a larger exposure of basaltic bedrock) has been suggested to warm the surface by, e.g., \citet{fairen2012} and \citet{mischna2013}.

With 3D model studies (e.g., \citealp{johnson2008}, \citealp{wordsworth2013}, \citealp{mischna2013}, \citealp{urata2013}), the problem of cold global mean surface temperatures could be addressed to some extent: They showed that even for mean surface temperatures below freezing, large areas of the Martian surface could remain much warmer, with annual means of 260-270\,K. In addition, 3D global and mesoscale models of the early Mars climate suggest that orography could be an important factor to drive precipitation. In a recent study, \citet{scanlon2013} show that orography-driven precipitation in the form of snowfall (of the order of about 10$^{-2}$-10$^{-1}$\,kg d$^{-1}$ m$^{-2}$) coincides roughly with the location of former rivers on early Mars.

\subsection{Background on Geomorphology: Martian Valley Networks }
\label{geo_back}

The term €œvalley networks denotes fluid-carved systems of incisions on planetary surfaces, interpreted to be former river valleys. Less degraded fluvial valleys may still possess a narrow interior channel along the valley bottom, which represents the riverbed itself (e.g., \citealp{jaumann2005}). Valley networks on Mars occur in two generic types, namely "dendritic" and "longitudinal". Each type implies a different hydrological regime. Dendritic patterns are interpreted to be indicative of precipitation-fed surface runoff due to their analogy to terrestrial features (e.g., \citealp{craddock2002}, \citealp{irwin2005}, \citealp{barnhart2009}, \citealp{ansan2013}). The surface runoff can be either caused by snowmelt or rain whereby recent studies emphasize that episodic snowmelt might be the most favorable process of water release (e.g., \citealp{forget2013}, \citealp{wordsworth2013}, \citealp{scanlon2013}). Longitudinal valleys may represent fluvial channels, but featuring only a few tributaries. 

Whereas some authors propose erosion by groundwater seepage (sapping) as the most plausible water release mechanism for these channels (e.g., \citealp{malin1999}, \citealp{goldspiel2000}, \citealp{harrison2005}, \citealp{jaumann2010}), others have demonstrated that sapping alone does not account for the erosion at analogous terrestrial channels and a significant contribution by overland runoff is required (e.g., \citealp{lamb2008}). Valley networks on Mars occur mostly in the heavily cratered southern highlands whereas some isolated fluvial channels have been observed along the flanks of volcanic edifices (e.g., \citealp{gulick1990}, \citealp{carr1995}, \citealp{fassett2006}, \citealp{hynek2010}).

Crater size-frequency analyses of valley network-incised regions show that fluvial activity peaks during the late Noachian and sharply decreases after the early Hesperian (e.g., \citealp{fassett2008,fassett2011}, \citealp{hoke2009}). Nevertheless, recent research has shown that aqueous surface processes continued even after the early Hesperian, though on a less intense level (e.g., \citealp{fassett2010}, \citealp{howard2011}, \citealp{hauber2013}, \citealp{parsons2013}, \citealp{hobley2014}). A recent study by \citet{buhler2014} suggests intermittent (not continous) fluvial activity of the order of 10$^{-3}$ of the available time to form the networks, based on complex transport and hydrological analyses.

\section{Tools and Methods}

\label{tools}

\subsection{Atmosphere}

We use atmospheric profiles of pressure $p$, temperature $T$ and water concentrations $c_{\rm{H2O}}$ from calculations presented in \citet{vparis2013marsn2}. These profiles were obtained with a 1D steady-state, radiative-convective atmospheric model which simulates globally averaged conditions. The model solves the radiative transfer equation and accounts for convective energy transport in the lower atmosphere by performing instantaneous convective adjustment to the (wet) adiabatic lapse rate. Further details can be found in \citet{vparis2013marsn2} and references therein. Model atmospheres were assumed to be composed of varying amounts of CO$_2$ (8 values between 0.02-3\,bar) and N$_2$ (5 values between 0-0.5\,bar). Solar irradiation was set to be consistent with Noachian conditions 3.8 billion years ago, i.e. 75\,\% of today's irradiation (e.g., \citealp{gough1981}).

\begin{figure}[h]
  \centering
  \includegraphics[width=200pt]{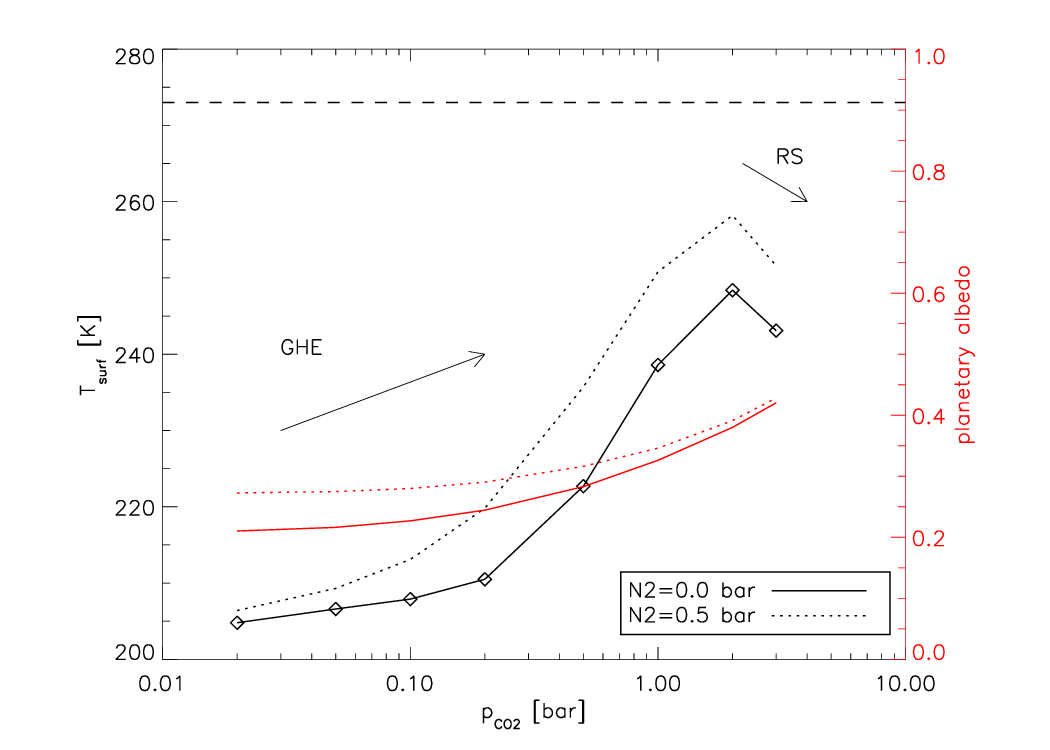}\\
  \caption{Surface temperatures (black) and planetary albedo (red) as a function of CO$_2$ partial pressure at two values of N$_2$ partial pressure (plain: 0\,bar, dotted: 0.5\,bar), values taken from simulations of \citet{vparis2013marsn2}. 273\,K indicated by horizontal dashed line. RS=Rayleigh scattering (cooling), GHE=Greenhouse effect (warming).}\label{ts}
\end{figure}

Figure \ref{ts} shows the calculated surface temperatures (black lines) and planetary albedos (red lines). The planetary albedo increases due to enhanced Rayleigh scattering which scales with atmospheric mass, hence also the higher albedo in the 0.5\,bar N$_2$ case. With increasing CO$_2$ partial pressures, both the zero N$_2$ case (plain line) and the 0.5\,bar N$_2$ case (dotted line) approximately converge to the same albedo since CO$_2$ is a much more efficient scatterer than N$_2$ \citep{vardavas1984} and dominates the scattering optical depth. Surface temperatures show a distinct maximum. This is because the surface temperature first increases when increasing CO$_2$ due to an enhanced greenhouse effect, reaching 248\,K at 2\,bar of CO$_2$ and zero N$_2$. Adding 0.5\,bar of N$_2$ increases the surface temperature by up to 12\,K, depending on CO$_2$ partial pressure. Above (2-3)\,bars of surface CO$_2$, the greenhouse effect saturates. An opposing, cooling effect via the strong Rayleigh scattering becomes important, and surface temperatures decrease again. This effect is known as the maximum greenhouse effect (e.g., \citealp{kasting1991}, \citealp{kasting1993}, \citealp{tian2010}, \citealp{vparis2013marsn2}).

Water concentration profiles as a function of altitude $z$ in \citet{vparis2013marsn2} were calculated from the following equation:
\begin{equation}\label{ch2o}
  c_{\rm{H2O}}(z)=\rm{RH}(z) \cdot \frac{p_{\rm{sat,H2O}}(T(z))}{p(z)}
\end{equation}

where $\rm{RH}$ is the Relative Humidity and $p_{\rm{sat,H2O}}(T(z))$ the temperature-dependent water saturation vapour pressure. RH is thus defined as the water vapour pressure relative to the saturation vapour pressure and is a measure of the amount of water able to be held in the gas-phase.  It depends on e.g. T-p conditions, and (sensitively) to the heterogenous surface loading (dust, aerosol) of the atmosphere (which is not known for early Mars). RH is therefore not well constrained in the early Martian atmosphere. Previous 1D studies usually assumed either a fully-saturated troposphere, i.e. RH=1, or 50\,\% saturation, i.e. RH=0.5 (e.g., \citealp{mischna2000}, \citealp{Cola2003}, \citealp{tian2010}). The calculations in \citet{vparis2013marsn2} also used RH=1. For 1D simulations of modern Earth or hypothetical terrestrial exoplanets, the RH profile is often based on the observed mean Earth profile of \citet{manabewetherald1967} (RH=MW, e.g., \citealp{Seg2003}, \citealp{grenf2007pss}, \citealp{wordsworth2010}).

From the calculated water profiles of Eq. \ref{ch2o}, it is possible to obtain the mean atmospheric water column ($C_{\rm{H2O}}$ in units of kg m$^{-2}$, see Figure 3 in \citealp{vparis2013marsn2}), via:

\begin{equation}\label{column}
  C_{\rm{H2O}}=\frac{m_{\rm{H2O}}}{g \mu}\cdot\sum_{i=1}^{N-1}c_{\rm{H2O,i}}(p_i-p_{\rm{i+1}})
\end{equation}

where $N$ is the number of atmospheric levels (here, $N$=52, $i$=1 at surface), $g$ the planetary gravity (for Mars, $g$=3.73\,m s$^{-2}$), $\mu$ the mean atmospheric weight, $m_{\rm{H2O}}$ the molecular weight of water and $c_{\rm{H2O,i}}$, $p_i$ water concentrations and pressure in level $i$.

In atmospheric 1D and 3D models, precipitation is calculated, e.g., by assigning precipitation efficiencies to cloud layers (e.g., \citealp{renno1994}), choosing a precipitation threshold for the water content (e.g., \citealp{wordsworth2013}) or simply assuming that clouds at the surface precipitate all super-saturated water (e.g., \citealp{segura2008}). In this work, we choose a different approach, as follows. The mean precipitation on Earth is $\approx$2.6\,mm d$^{-1}$, i.e. 2.6\,kg m$^{-2}$ d$^{-1}$ ( e.g., \citealp{xie1997}, \citealp{mitchell2005}, \citealp{adler2012}). Taking a mean Earth water column of $C_{\rm{H2O,Earth}}$=19.5\,kg m$^{-2}$ (see e.g., modern Earth atmospheric simulations by \citealp{grenf2007pss}), this amounts to a daily precipitation rate, $pr$=13.3\% of the total water column. In this work we apply this value of $pr$ (which is clearly a source of uncertainty, see discussion below) to the early Mars scenarios (see Fig. \ref{ts}) to calculate precipitation rates $P$:

 \begin{equation}\label{prec}
   P=pr \cdot C_{\rm{H2O}}
 \end{equation}

 For example, \citet{vparis2013marsn2} found, at a CO$_2$ partial pressure of 2\,bar and 0.5\,bar of N$_2$, a water column of 5.2 kg m$^{-2}$ (see their Fig. 3). The corresponding mean precipitation would then be 0.7 mm d$^{-1}$, i.e. around 26\% of the mean Earth value.

To explore uncertainties in this simple, first-order approach, we perform the following sensitivity studies regarding three important parameters for the calculation of precipitation, i.e. the assumed percentage precipitation $pr$, the RH profile and the surface temperature:

\textbf{Atmospheric Precipitation}

\begin{figure}[h]
  \centering
  \includegraphics[width=300pt]{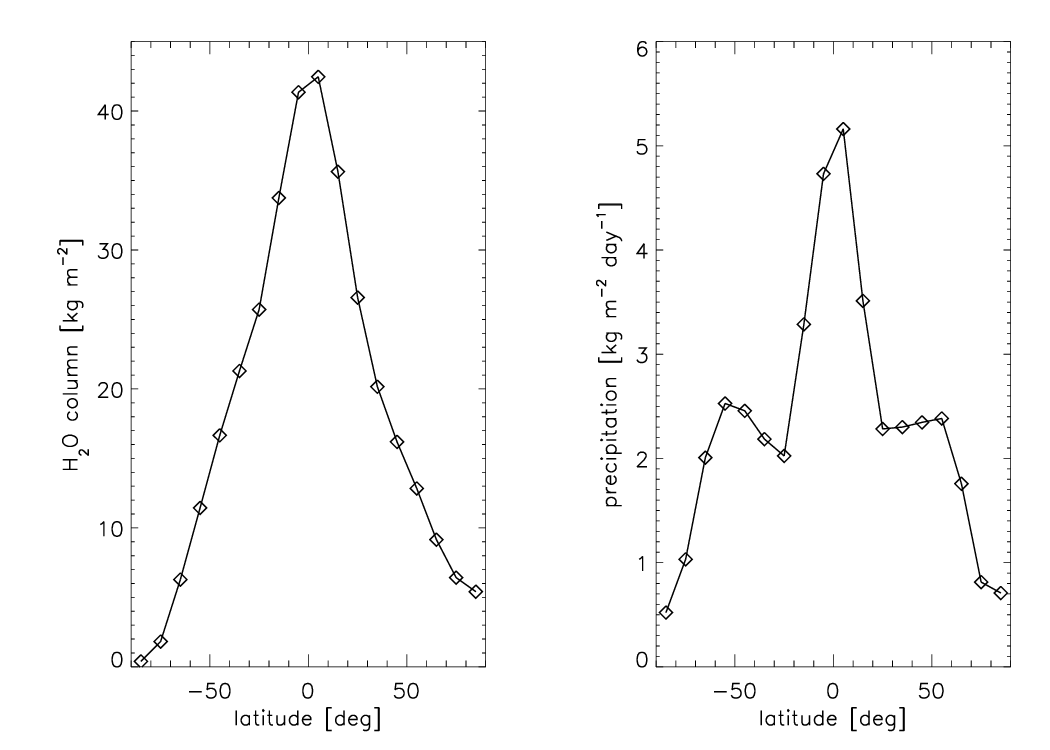}\\
  \caption{Latitudinal gradient of water column (left panel) and precipitation (right panel). 10$^{\circ}$-averaged annual mean for year 2013, based on the NCEP/NCAR project \citep{kalnay1996}. Data set available through http://www.esrl.noaa.gov/psd/cgi-bin/data/timeseries/timeseries1.pl}\label{climatology}
\end{figure}

\begin{figure}[h]
  \centering
  \includegraphics[width=300pt]{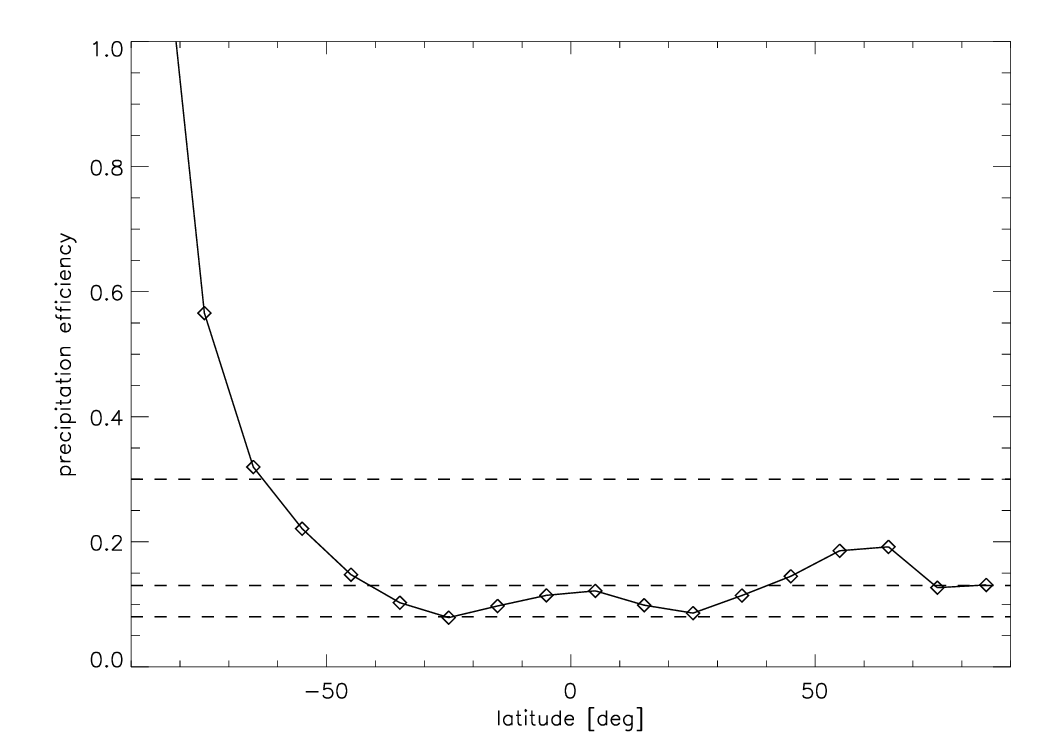}\\
  \caption{Latitudinal gradient of precipitation rate $pr$, based on Fig. \ref{climatology}. }\label{preceff}
\end{figure}

We vary $pr$ in Eq. \ref{prec} assumed for early Mars as follows. On Earth (mean $pr$=13.3\%), there is a latitudinal gradient in $pr$, due to global circulation and the land-ocean distribution. Figure \ref{climatology} shows the latitudinal gradient of the annual means of water column and precipitation. The values are based on monthly averages for the year 2013, accessed through the NCEP/NCAR \citep{kalnay1996} website http://www.esrl.noaa.gov/psd/cgi-bin/data/timeseries/timeseries1.pl. Figure \ref{preceff} then shows the associated value of $pr$, using eq. \ref{prec}. For the most part (except south polar regions, where water column measurements are very uncertain), $pr$ remains between about 8-30\,\%. Hence, we varied $pr$ based on the extremes found on modern Earth, i.e. $pr$=8, 13.3 and 30\%. Idealized 3D simulations by \citet{ogorman2008} found that the atmospheric water residence time $\tau_{w}=1/pr$ depends on surface temperature and actually increases for warmer climates, broadly consistent with modern Earth observations (see Fig. \ref{preceff}). The actual amount of precipitation depends on the amount of available water in the atmospheric column as well as the precipitation efficiency $pr$, which can both be calculated consistently in a 3D model. However, in this work, we only calculate the water column as a function of atmospheric composition and vary $pr$ within a reasonable range.

How accurate is this approach? Using 3D GCM simulations of early Mars presented by \citet{wordsworth2013}, we estimated their $pr$ value to be of the order of 20-50\% (their Fig. 10 and Table 2), comparable to polar values on Earth (see Fig. \ref{preceff}). This estimate of $pr$ is obtained as follows: On summing (by eye) the panels describing seasonal snowfall in Fig. 10 of \citet{wordsworth2013} and using their stated water column of the 1\,bar simulation (0.07\,kg\,m$^{-2}$, see their Table 2), we found a range for $pr$ of 20-50\,\%. On early Mars, clearly several factors such as the Martian dichotomy will influence global convection, hence the strength of the Hadley cell.  Nevertheless these factors are not well defined so our assumption based on the Earth is a reasonable first estimate.

\textbf{	Atmospheric Relative Humidity }

\begin{figure}[h]
  \centering
  \includegraphics[width=250pt]{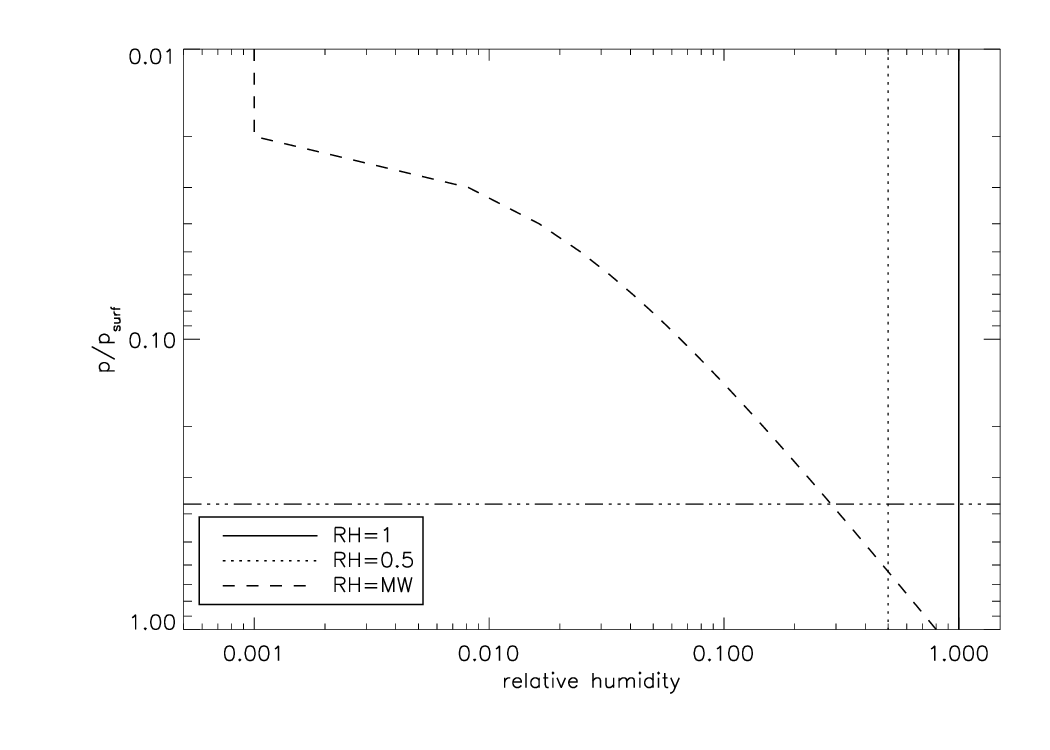}\\
  \caption{Relative humidity profiles used in this work. Lowermost atmospheric scale height indicated by horizontal triple-dot-dashed line.}\label{rhprof}
\end{figure}

To investigate uncertainties in early Mars' RH profiles (see above), we introduced two new RH test cases, namely (i) RH=0.5 and (ii) RH=MW, in addition to the original RH=1 from \citet{vparis2013marsn2}. These three profiles are shown in Fig. \ref{rhprof}. Note that although the MW profile drops sharply with altitude, within the lowermost atmospheric scale height ($\frac{p}{p_{\rm{surf}}}=0.36$, indicated by horizontal line) where most of the water resides, the three RH profiles are rather similar.

\begin{figure}[h]
  \centering
  \includegraphics[width=250pt]{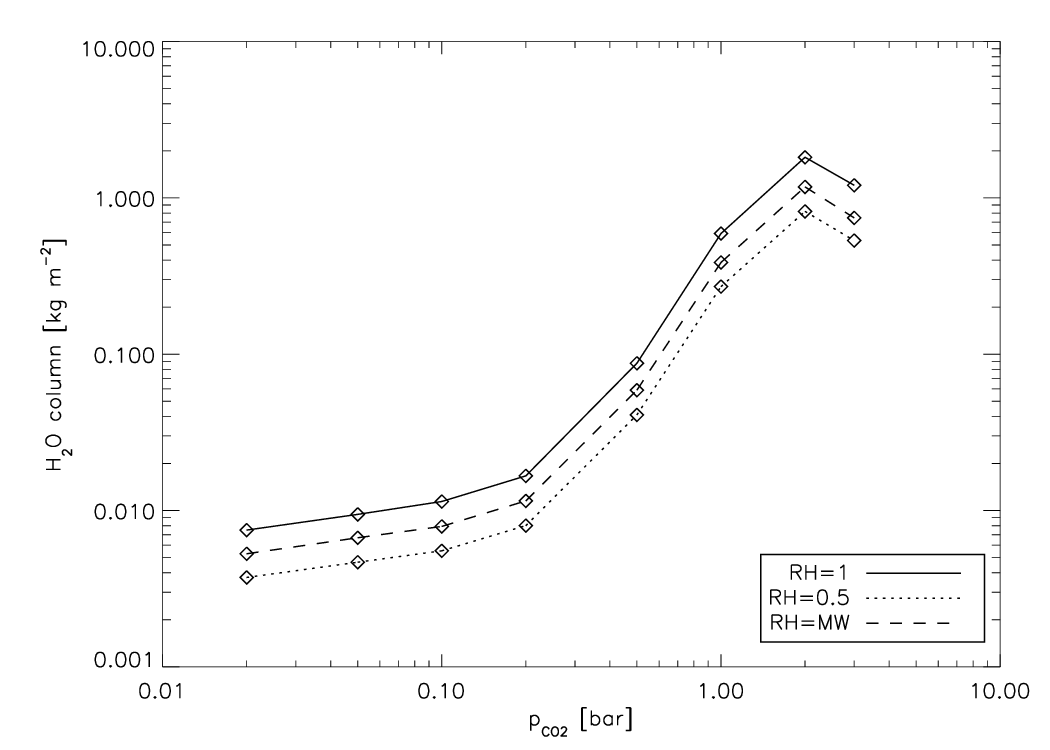}\\
  \caption{H$_2$O column as a function of CO$_2$ pressure: Influence of relative humidity profiles. Zero N$_2$.}\label{cumul}
\end{figure}

We repeated the zero-N$_2$ simulations with the atmospheric model of \citet{vparis2013marsn2}, using the new RH profiles (i.e., RH=0.5 and RH=MW). The effect on surface temperature was on the order of a few K, comparable to 1D results obtained by \citet{Cola2003}. Note that 3D simulations performed by \citet{wordsworth2013} find a somewhat larger effect of up to 20\,K. Water columns decreased by about 30-50\,\% when compared to the RH=1 case, as shown in Fig. \ref{cumul}.

\textbf{Surface temperature} 

Here, we estimated the effect of a (local) surface temperature increase $\Delta T$ (with respect to the zero N$_2$ scenarios) upon precipitation (see Fig. \ref{deltat_contour} and discussion there). The surface temperature and the temperature profile are very important factors when calculating precipitation rates because the water vapour saturation pressure $p_{\rm{sat,H2O}}$, hence the water concentration profile (see Eq. \ref{ch2o}), depend strongly on temperature. We add $\Delta T$ to both the surface temperature and the temperature profile from \citet{vparis2013marsn2}. The water profile and the resulting column are then re-calculated with Eqs. \ref{ch2o} and \ref{column}. This simplifying approach is justified since most of the water column resides near the surface where the temperature lapse rate is approximately constant with altitude, at about 3-4\,K km$^{-1}$, which is close to the dry adiabatic value of 4.3\,K km$^{-1}$.

\textbf{Global mean vs local precipitation} 

\begin{figure}[h]
  \centering
  \includegraphics[width=250pt]{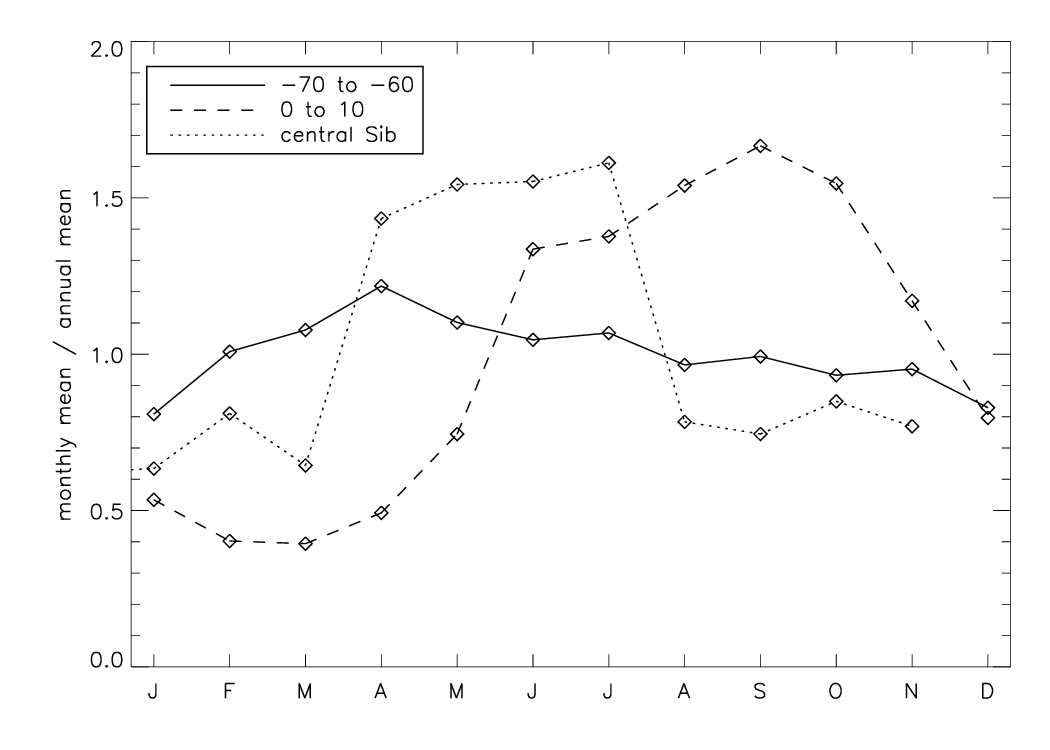}\\
  \caption{Variation of the monthly mean precipitation around the annual mean for different locations on Earth (zonal means at given latitude and in Central Siberia). Data set available through http://www.esrl.noaa.gov/psd/cgi-bin/data/timeseries/timeseries1.pl}\label{xval}
\end{figure}

There exists not only a latitudinal gradient in mean precipitation (Fig. \ref{climatology}), but also a monthly variation, as shown in Fig. \ref{xval}. Therefore, at any given location and given time, the actual local precipitation $P_{\rm{loc}}$ might not be well represented by the global  mean precipitation $P_{\rm{glob}}$. Rather, in an extremely simplified approach, one might relate these through 

\begin{equation}
\label{enhance}
P_{\rm{loc}}=x \cdot P_{\rm{glob}}
\end{equation}

where $x$ must be determined by temporally and spatially resolved calculations. However, as indicated by Figs. \ref{climatology} and \ref{xval}, on spatial scales resolved by present-day 3D GCM simulations, $x$ is probably not larger than about 3. Mesoscale 3D models of early Mars have shown that orography-driven precipitation can be much stronger (up to an order of magnitude) than synoptic-scale precipitation \citep{scanlon2013}, suggesting values of up to $x$=10 for increasingly finer spatial resolutions.

In a last parameter variation, we therefore estimated an approximate value for $x$ in order to obtain an agreement between atmospheric model and geomorphology data.

\subsection{Geological Constraints}

\subsubsection{Methods}

Most valley networks represent relics of ancient fluvial activity on early Mars near the Noachian-Hesperian boundary. Thus, it is not possible to measure directly discharge rates $Q$ (in units of m$^{3}$s$^{-1}$). They are instead derived from parameters related to the valley system'€™s morphometric properties, channel width $W_C$, depth $D$, and the flow velocity $v$:

\begin{equation}\label{qdef}
  Q=W_C\cdot D \cdot v
\end{equation}

The depth $D$ of the Martian riverbed is usually not available. It is in most cases too small to be resolved in the Digital Terrain Models (DTM) derived from, e.g. Mars Express's HRSC (High Resolution Stereo Camera, \citealp{jaumann2007}, \citealp{gwinner2010}) and Mars Global Surveyor'€™s MOLA (Mars Orbiter Laser Altimeter, \citealp{smith2001}). Furthermore, the flow velocity $v$ is also unknown. Therefore, $Q$ can only be derived based on the channel width $W_C$, which is in principle observable. The lack of reliable information on the stability of the channel banks complicates accurate measurements of $W_C$. However, empirical correlations can be used to estimate discharge rates from the width of the channels. On Earth, for channels in the Missouri river area, \citet{osterkamp1982} use an empirical power law equation ($Q=f\cdot W_C^e$) to obtain mean discharge rates:

\begin{equation}\label{qmean}
  Q_{\rm{Earth,mean}}=f_{\rm{Earth}} \cdot W_C^{e_{\rm{Earth}}}=0.027 \cdot W_C^{1.71}
\end{equation}

The main assumption in this work and others (e.g., \citealp{irwin2005}) is that $Q$ scales with the same power law, meaning the exponent $e_{\rm{Earth}}$ does not change for Martian channels (i.e., $e_{\rm{Earth}}$=$e_{\rm{Mars}}$). Only the scaling factor $f_{\rm{Mars}}$ will be adjusted to Martian gravity based on the value of $f_{\rm{Earth}}$. To obtain a conservative estimate of the channel width, we follow the approach of \citet{irwin2005}. We use narrow, straight channel sections to measure channel widths. It is assumed that channel bank-to-bank widths are less modified in such areas. Thus, this approach helps minimizing the impact of subsequent mass movements that could heavily modify channel bank-to-bank widths.

To calculate $f_{\rm{Mars}}$, we follow the approach of \citet{irwin2005} and \citet{moore2003}. They assumed that the channel width $W$ scales with relative gravity $g$ (=0.38 for Mars/Earth system) as g$^{-0.23}$. For the same unit discharge $Q$, since $e_{\rm{Earth}}$=$e_{\rm{Mars}}$, we have then:

\begin{equation}\label{e_scale}
f_{\rm{Mars}}=f_{\rm{Earth}} \left(\left(\frac{g_{\rm{Mars}}}{g_{\rm{Earth}}}\right)^{-0.23}\right)^{-e_{\rm{Mars}}}=0.018
\end{equation}

Inserting into eq. \ref{qmean} leads to the following empirical relation between $W$ and $Q$ on Mars

\begin{equation}\label{qmeanmars}
Q_{\rm{Mars,mean}}=0.018 W_C^{1.71}
\end{equation}

Local variations of $W$ by later erosion and modification processes result in a possible error of a factor of 2; a further factor of 3 may arise via the unknown stability of the banks and is also inherent in the empirical equation itself \citep{irwin2005}. Thus, the results for $Q$ are conservative estimates, but are most likely accurate to within an order of magnitude. Once the discharges have been derived, the runoff rate $R$ can then be estimated:

\begin{equation}\label{runoff}
  R=\frac{Q}{C}
\end{equation}

where $Q$ is the discharge rate of a valley network at its outlet and $C$ the respective catchment area. This results in a water equivalent layer from a few millimeters up to several centimeters thick per d.

\subsubsection{Data set}

\begin{figure}[h]
  \centering
  \includegraphics[width=350pt]{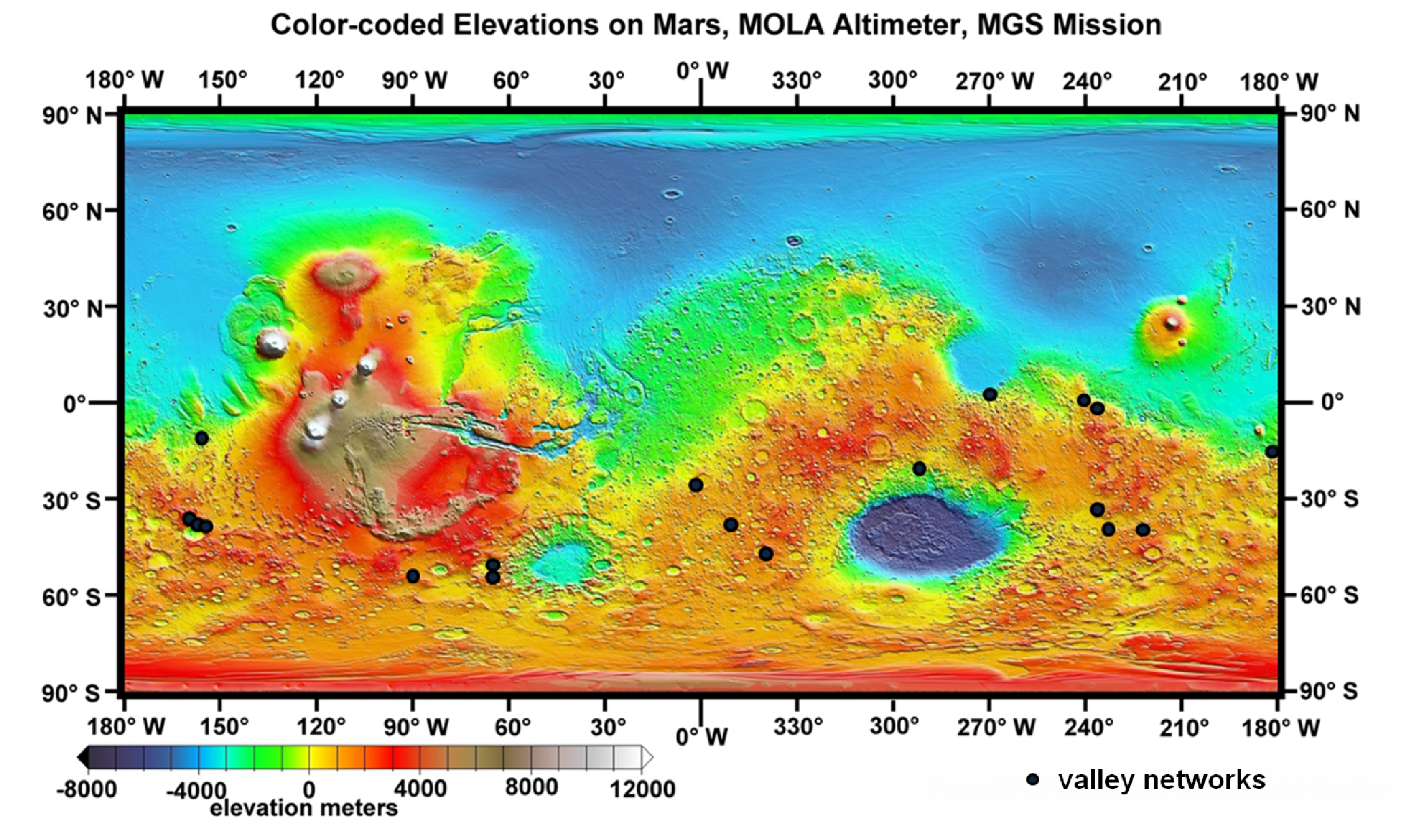}\\
  \caption{Distribution of the 18 valley networks (black dots) used in this work.}\label{distri}
\end{figure}

In total, we used runoff values of 18 valley networks in our analysis, as presented in Table \ref{tabledata}.  Their distribution is indicated in Fig. \ref{distri}. It is generally accepted that surface runoff occurred mainly at the transition between Noachian/Hesperian 3.7-3.8 billion years ago and earlier (e.g., \citealp{fassett2008}, \citealp{hynek2010}). At this time, there was a major decrease in runoff intensity on the Martian surface \citep{carr2010}. Hence the data set is assumed to be consistent in time with the atmospheric model simulations.

\begin{table*}
%  \centering
  \caption{Valley networks used in this work (measured quantities in bold).}\label{tabledata}
%  \begin{tabular}{lp{3cm}lp{3cm}cp{0.5cm}cp{0.5cm}cp{0.5cm}cp{3cm}cp{3cm}}
\noindent\makebox[\textwidth]{%
  \begin{tabular}{p{3cm}p{3cm}p{0.8cm}p{0.8cm}p{0.8cm}p{1.2cm}p{1.4cm}p{1.2cm}p{1.4cm}}
     \hline
   \hline
    Identifier & location&$W_V$ [m] &$W_C$ [m] & $C$ [km$^2$]& $Q_{\rm{mean}}$ [m$^3$ s$^{-1}$]& $R_{\rm{mean}}$ [mm d$^{-1}$]&$Q_{\rm{peak}}$ [m$^3$ s$^{-1}$]&$R_{\rm{peak}}$ [mm d$^{-1}$]\\
    \hline
    \hline
   8604-1               &   38$^{\circ}$15' S/ 203$^{\circ}$ E&-                  & \textbf{250}       &\textbf{1840} & 226.8  & 10.65  &1179.2& 55.3    \\
   8604-2               &   38$^{\circ}$27' S/ 204$^{\circ}$ E&-                  & \textbf{220}       &\textbf{2970} & 182.3  & 5.30    &1008.9&  29.3  \\
   8604-3               &   38$^{\circ}$30' S/ 204$^{\circ}$5' E&-                  & \textbf{255}       &\textbf{2230} & 234.6  & 9.09      &1208.1&  46.8\\
   H1226\_0000\_ND3               &   1$^{\circ}$52' N/ 89$^{\circ}$25' E&\textbf{1770}                  & 247      &\textbf{12840} & 222.2  & 1.49 &1162.0& 7.8       \\
   H5212\_0000\_ND3               &   1$^{\circ}$49' N/ 121$^{\circ}$16' E&\textbf{5850}                  & 819      &\textbf{21240} & 1725.7  & 7.02  &5015.7&   20.4   \\
   H2081\_0000\_ND3               &   0$^{\circ}$15' N/ 124$^{\circ}$12' E&\textbf{2070}                  & 289      &\textbf{4240} & 290.6  & 5.92      &1407.4&  28.6\\
   H0430\_0000\_ND3               &   36$^{\circ}$57' S/ 7$^{\circ}$56' E&\textbf{480}                  & 67      &\textbf{300} & 23.8  & 6.87       &236.5& 68.1 \\
   H2181\_0001\_ND3               &   45$^{\circ}$29' S/ 17$^{\circ}$16' E&\textbf{1430}                  & 200      &\textbf{3030} & 154.8  & 4.41  &898.2& 25.6     \\
   H5168\_0001\_ND3-2            &   10$^{\circ}$47' S/ 156$^{\circ}$45' W&\textbf{1180}                  & 165      &\textbf{3530} & 111.4  & 2.72    &710.3&    17.3\\
   H2689\_0001\_ND3               &   34$^{\circ}$40' S/ 146$^{\circ}$04' E&\textbf{2120}                  & 296      &\textbf{9070} & 302.8  & 2.88      &1449.1& 13.8 \\
   H7213\_0000\_ND3               &   12$^{\circ}$21' S/ 177$^{\circ}$58' W&\textbf{2040}                  & 285      &\textbf{14120} & 283.8  & 1.73      &1383.6& 8.4 \\
   H2459\_0009\_ND3               &   17$^{\circ}$6' S/ 65$^{\circ}$42' E&\textbf{2790}                  & 390      &\textbf{15950} & 485.2  & 2.62      &2028.7&  10.9\\
   H2347\_0000\_ND3               &   52$^{\circ}$49' S/ 90$^{\circ}$49' W&\textbf{1880}                  & 263      &\textbf{8400} & 247.3  & 2.54      &1254.5& 12.9 \\
   H2475\_0000\_ND3-1            &   49$^{\circ}$19' S/ 65$^{\circ}$28' W& -                & \textbf{160}      &\textbf{6040} & 105.7  & 1.51      &684.1& 9.7 \\
   H2475\_0000\_ND3-3            &   52$^{\circ}$26' S/ 65$^{\circ}$51' W& \textbf{1030}                & 144      &\textbf{2430} & 88.3  & 3.14     &601.6& 21.3  \\
   H4290\_0000\_ND3               &   35$^{\circ}$3' S/ 132$^{\circ}$ E& -               & \textbf{270}      &\textbf{7170} & 258.7  & 3.11      &1295.3& 15.6 \\
   H2539\_0000\_ND3               &   27$^{\circ}$16' S/ 128$^{\circ}$10' E& \textbf{1600}               & 224      &\textbf{1720} & 188.0  & 9.44  &1031.4& 51.8      \\
   H6438\_0000\_ND3               &   24$^{\circ}$54' S/ 3$^{\circ}$26' W& \textbf{2880}               & 403      &\textbf{18230} & 513.2  & 2.43      &2111.5& 10.0 \\
     \hline
          \end{tabular}
}
\end{table*}

Their channel widths and the corresponding discharge rates have been determined. Where available, direct measurements of $W_C$ have been used to calculate $Q$ via Eq. \ref{qmeanmars}. However, most of these valleys do not show any channels (because they are completely covered with sediment or destroyed by erosion). Hence, channel widths $W_C$ for Eq. \ref{qmeanmars} have been deduced from the measured valley width $W_V$ via a simple scaling suggested by \citet{penido2013}:

\begin{equation}\label{valleyw}
  W_C=0.14 \cdot W_V
\end{equation}

Subsequently, the sizes of the respective catchment areas $C$ were determined using topographic data based by HRSC and MOLA DTMs (e.g., \citealp{gwinner2010}, \citealp{smith2001}).

The runoff rates were calculated from Eq. \ref{runoff}, resulting in a mean runoff of 1.5-10.65\,mm d$^{-1}$, depending on valley location. The median value of our dataset is 3.14\,mm d$^{-1}$, relatively close to the mean Earth precipitation of 2.6\,mm d$^{-1}$. As is apparent from Fig. \ref{climatology} and Table \ref{tabledata}, quite a few valley networks require precipitation rates close to or higher than typical tropical values found on Earth. Overall, therefore, these runoff rates would suggest a very strong, global mean hydrological cycle on early Mars, if the networks indeed formed throughout extended periods of warm climate.

\begin{table*}
  \centering
  \caption{Comparison of peak discharge and runoff rates.}\label{comparepeak}
  \begin{tabular}{lccl}
     \hline
   \hline
    Region/name &  $Q_{\rm{peak}}$ [m$^3$ s$^{-1}$]& $R_{\rm{peak}}$ [cm d$^{-1}$]& references\\
    \hline
    \hline
  
   global distribution (see Fig. \ref{distri})               &  236-5015       & 0.8-6.8 &      this work  \\
   global distribution            &  300-5800       & 0.1-6 &      \citet{irwin2005}  \\
   Terra Sabaea, Arabia Terra, Meridiani Plan.             &  7,000-70,000     & 0.4-63 &      \citet{hoke2011} \\
  % Eberswalde, Isidis              &  700-1300     & - &      \citet{howard2005} \\
  % Libya Montes               & 4800     & - &      \citet{jaumann2005} \\
 % Parana Basin (Samara Vallis)             & 100-4600     & - &      \citet{barnhart2009} \\
% Eberswalde            & 500-1000     & - &      \citet{mangold2012} \\
% Newton            & -     & 1-10 &      \citet{parsons2013} \\
     \hline
          \end{tabular}
\end{table*}

\subsubsection{Comparison with previous work}

To compare the runoff rates for our networks with rates previously published in the literature, we need to calculate peak instead of mean runoff rates (also shown in Table \ref{tabledata}). Note that when comparing to a global mean atmospheric model, as is the aim in this work, we should use mean runoff rates. In a similar reasoning as for deriving eq. \ref{qmeanmars}, based on the peak discharge approximation from \citet{osterkamp1982} we obtain 

\begin{equation}\label{qpeakmars}
Q_{\rm{Mars,peak}}=1.4 W_C^{1.22}
\end{equation}

Peak runoff rates from our 18 valley networks are between 0.8\,cm d$^{-1}$ and 6.8\,cm d$^{-1}$. \citet{irwin2005} find runoff rates of 0.1\,cm d$^{-1}$ to 6\,cm d$^{-1}$.  In contrast to our study, \citet{hoke2011} investigated larger valley systems, which results in runoff rates of 0.4\,cm d$^{-1}$ to 63\,cm d$^{-1}$.  In general, our results are in good agreement with the measurements of other investigations of Martian peak runoff rates (see Table \ref{comparepeak}). In a very recent work, \citet{palucis2014} estimated discharge (not runoff) from a channel in Gale Crater (average channel width 27\,m). They found values ranging from 3.7-6.5\,m$^3$s$^{-1}$ when assuming a shallow channel and up to 117-207\,m$^3$s$^{-1}$ for deeper channels. This compares reasonably well to estimates from our simple equations \ref{qmeanmars} (5\,m$^3$s$^{-1}$) and \ref{qpeakmars} (78\,m$^3$s$^{-1}$) when using the stated channel width of 27\,m.

\section{Results}

\label{results}

\begin{figure}[h]
  \centering
  \includegraphics[width=200pt]{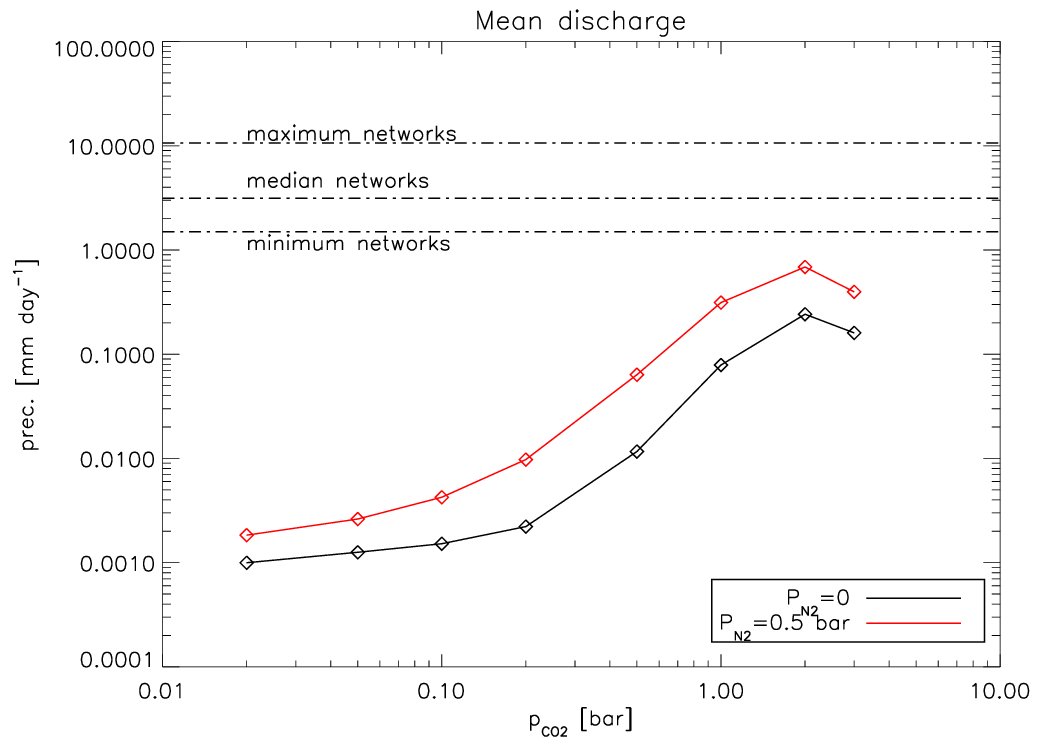}\\
  \caption{Total precipitation (\,mm d$^{-1}$) versus surface CO$_2$ pressure output from the atmospheric column model (\citealp{vparis2013marsn2}, RH=1) assuming $pr$=13.3\%. Black: No N$_2$, red: 0.5\,bar N$_2$. Mean discharge from Eq. \ref{qmeanmars}.}\label{comp}
\end{figure}

Figure \ref{comp} shows our precipitation rates as a function of CO$_2$ partial pressure as inferred from the atmospheric model for the scenarios discussed. Also shown are derived mean runoff rates (horizontal lines) from valley network data (minimum, median and maximum). We show atmospheric scenarios with zero and 0.5\,bar N$_2$. The atmospheric model results display a maximum in precipitation rate, as could be expected from Fig. \ref{ts}.

Figure \ref{comp} suggests that the calculated mean precipitation rates are lower by more than an order of magnitude compared to the median mean runoff value derived from the network data. Even the minimum runoff from the valley networks still is roughly 7 times higher than the maximum value calculated from the zero N$_2$ case.

\begin{figure}[h]
  \centering
  \includegraphics[width=200pt]{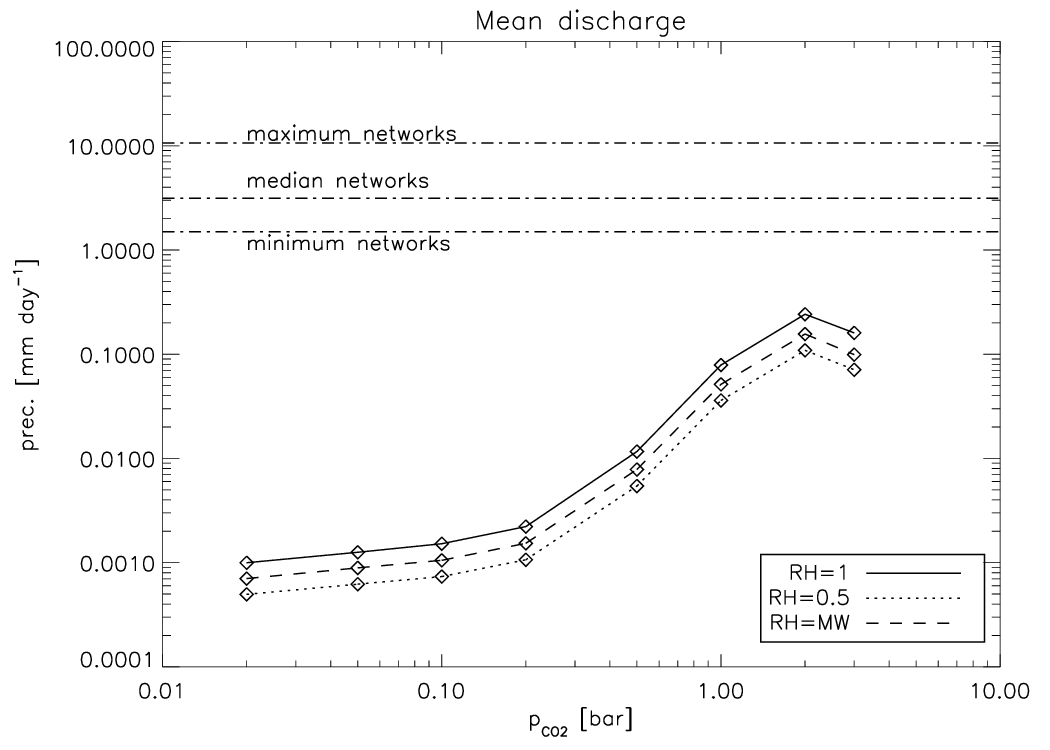}\\
  \caption{Total precipitation (mm d$^{-1}$) versus surface CO$_2$ pressure output from the atmospheric column model (\citealp{vparis2013marsn2}, zero N$_2$, $pr$=13.3\%). Effect of varying RH. Plain line: RH=1, dotted: RH=0.5, dashed: RH=MW.}\label{rheffect}
\end{figure}

Figure \ref{rheffect} shows the effect of adopting different relative humidity profiles on the calculated precipitation rate. As expected, since most of the water column lies in the lowermost atmospheric layers, the effect is rather small, even when reducing the RH from RH=1 to the MW RH profile, although it amounts to roughly a factor of 2 decrease compared to the fully saturated RH=1 case.

\begin{figure}[h]
  \centering
  \includegraphics[width=200pt]{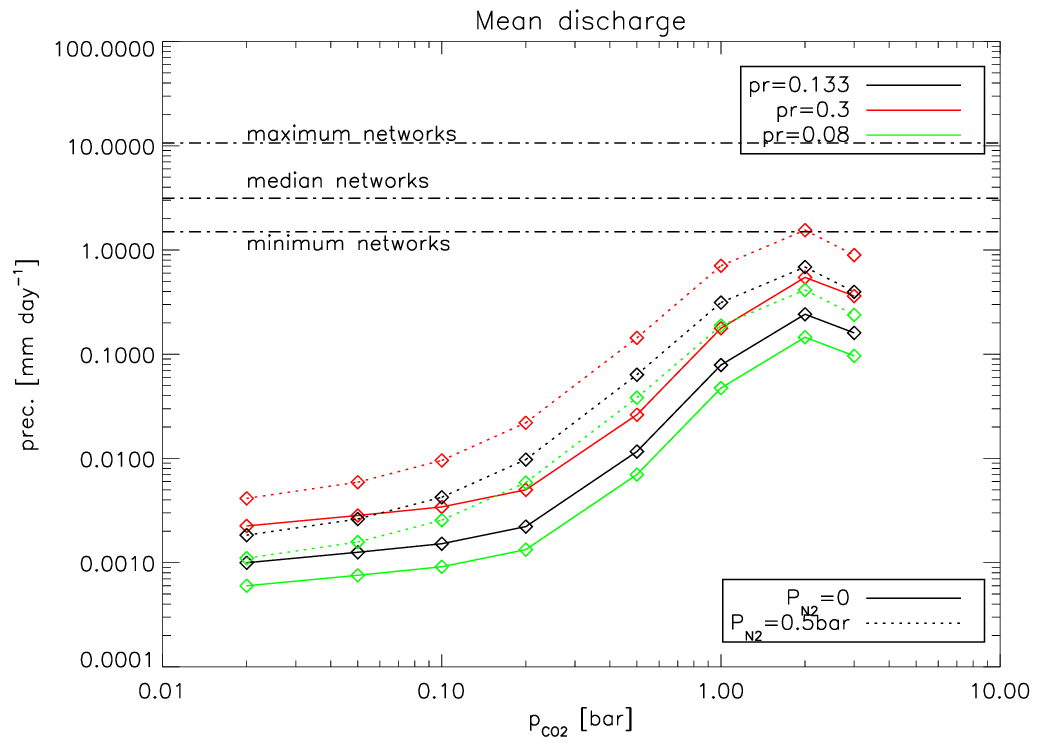}\\
  \caption{Total precipitation (mm d$^{-1}$) versus surface CO$_2$ pressure output from the atmospheric column model (\citealp{vparis2013marsn2}, RH=1, plain: zero N$_2$, dotted: 0.5\,bar N$_2$). Effect of varying percentage precipitation. Black: $pr$=13.3\%, red: $pr$=30\%, green: $pr$=8\%.}\label{preffect}
\end{figure}

Figure \ref{preffect} shows the influence of varying the adopted $pr$ value on precipitation rates. For the lower (Earth) values of $pr=8$\% the disagreement between atmospheric and geological data becomes stronger, as expected, i.e. a significant additional disagreement of a factor of about three between atmospheric and geological data becomes apparent. Only in the most favorable case (high N$_2$, high $pr$) can the atmospheric model reproduce roughly the lowest precipitation value inferred from the valley networks.

\begin{figure}[h]
  \centering
  \includegraphics[width=200pt]{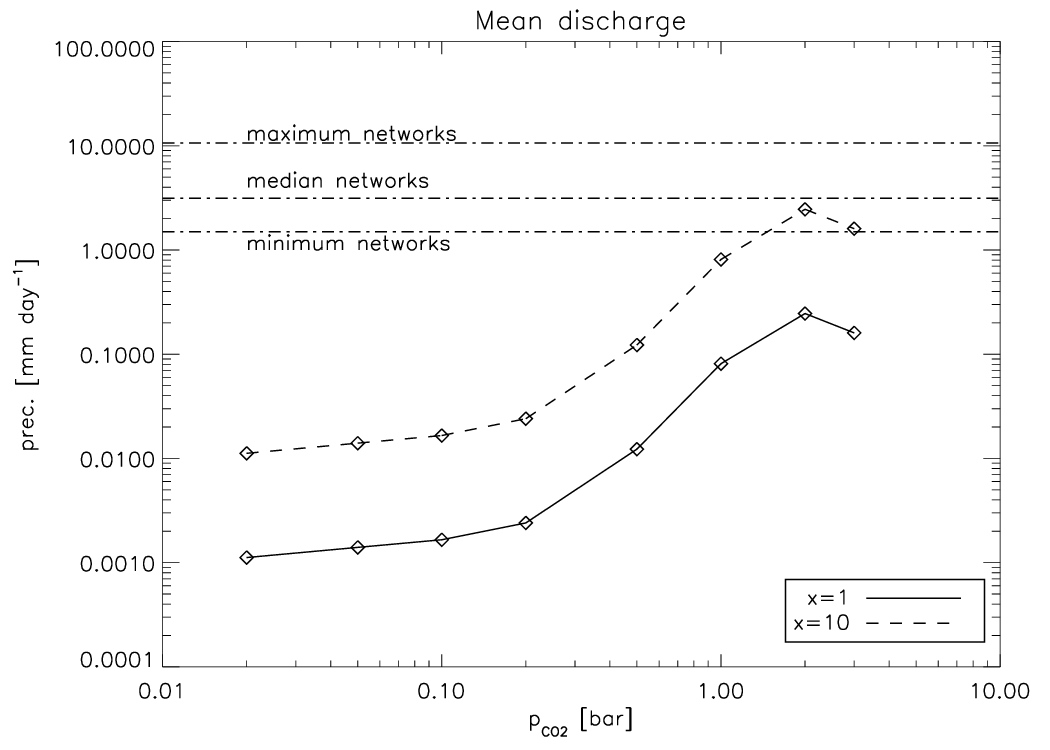}\\
  \caption{Most favored estimates of precipitation: RH=0.5, $pr$=0.3, for different values of $x$.}\label{realistic}
\end{figure}

Figure \ref{realistic} shows our favored estimates of early Mars precipitation, with RH=MW and $pr$=0.3. The use of RH=1 was justified by the absence of any realistic information on the relative humidity distribution on early Mars. Results by \citet{wordsworth2013} suggest that the atmosphere was most likely significantly drier, therefore we choose RH=0.5 which yields our driest runs (see Fig. \ref{cumul}). The favored estimate of early Mars precipitation is 0.3, which is closer to Earth polar values (see Fig. \ref{preceff}) and estimates from \citet{wordsworth2013}. We also show precipitation estimates with $x$=10 (taken from \citealp{scanlon2013}, see eq. \ref{enhance}).

It is clearly seen that in this case, the disagreement between global mean atmospheric and valley network data remains substantial. However, for large values of $x$, estimated amounts of precipitated water become somewhat closer to each other. Still, it is unclear which mechanisms would drive $x$ to the needed values of $x>$10-20 (or even higher at lower CO$_2$ pressures). This remains a question for future work.

\section{Discussion}

\label{disc}
\subsection{Runoff from rain or snowmelt?}

The aim of this work has been to compare precipitation rates calculated with a 1D atmospheric model to runoff rates inferred from geomorphological data. For this comparison to be meaningful, one critical condition must be fulfilled, i.e. precipitation $P$ is the source for runoff $Q$.

As is apparent from Fig. \ref{ts}, calculated global mean surface temperatures are far below freezing, hence $P$ would mainly occur as snowfall than as rain \citep{wordsworth2013} and therefore would not be linked directly to $Q$. Numerous studies suggest a possibly significant contribution by snowmelt (e.g., \citealp{clow1987}, \citealp{kite2013}, \citealp{wordsworth2013}, \citealp{palucis2014}), when orbital elements are suitable. Hence, snowmelt rather than precipitation may have been the dominant factor in producing liquid water available for runoff. Snow melt is a highly time-dependent process, and will concentrate discharge in river channels over short periods of time in spring and early summer (or whenever orbital elements favor local melt conditions), whereas discharge is much lower in the rest of the year (e.g., \citealp{bavay2009}). On Earth, runoff over the majority of the land area in the northern hemisphere is seasonally variable \citep{weingartner2013} and dominated by snowmelt (e.g., \citealp{ferguson1999}, \citealp{barnett2005}). This effect is particularly pronounced in arctic regions (e.g., \citealp{woo2012}), which may be considered terrestrial climatic analogs to early Martian environments. The hydrograph of arctic rivers displays a peak discharge during snowmelt season (e.g., \citealp{woo1986}, \citealp{boggild1999}), and it is this peak discharge which would be responsible for the channel-forming flood. However, we lack the knowledge as to how channel dimensions controlled by snowmelt and associated peak discharges can be used to infer annual mean precipitation rates on Mars.

In summary, the condition of liquid-water precipitation is unlikely to be fulfilled. Therefore, besides the obvious discrepancy in the amount of precipitated water (e.g., Fig. \ref{comp}), the fact remains that early Mars was most likely too cold to sustain network formation by continued liquid-water precipitation.

\subsection{Uncertainties in surface temperature calculations}

\begin{figure}[h]
  \centering
  \includegraphics[width=400pt]{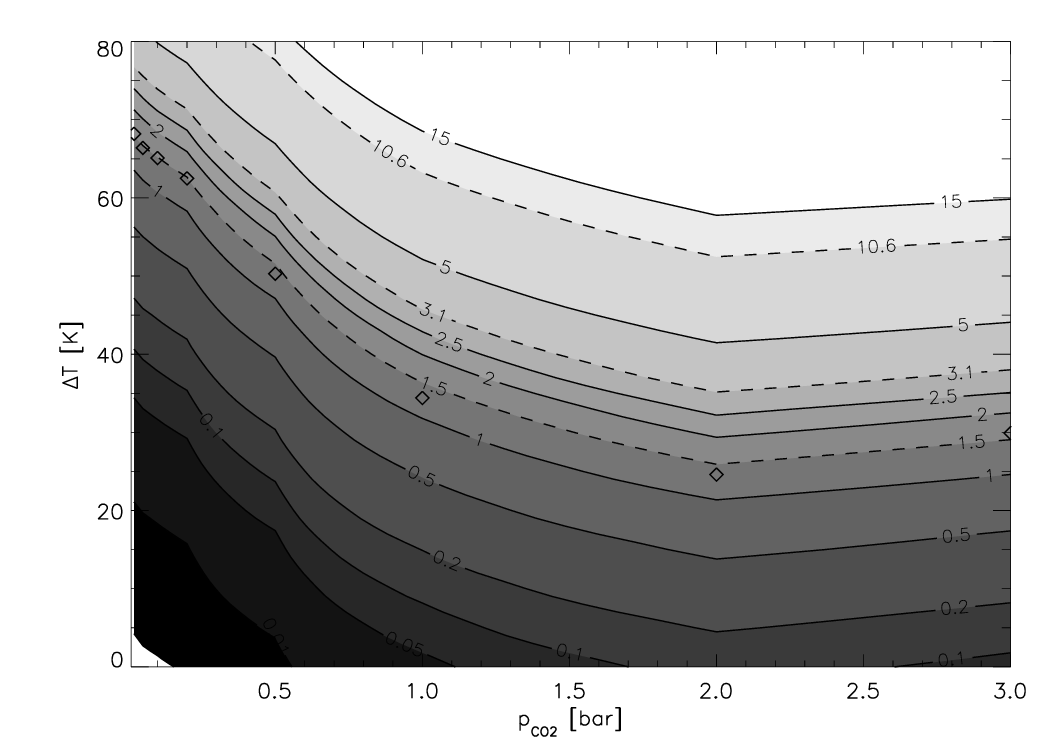}\\
  \caption{Mean precipitation (in mm d$^{-1}$) as a function of $\Delta$T and CO$_2$ partial pressure (zero N$_2$, RH=1, $pr$=13.3\%). Dashed lines represent minimum, mean and maximum runoff as inferred from the network data. $T$=273\,K indicated by $\diamond$ (for lower temperatures, precipitation would be in the form of snowfall). Note that the contribution of latent heat release due to evaporation is neglected, which would limit precipitation to about 1.8\,mm\,day$^{-1}$ (see Sect. \ref{limit_evap}).}\label{deltat_contour}
\end{figure}

Figure \ref{deltat_contour} shows the mean precipitation as a function of CO$_2$ partial pressure  and $\Delta T$. It is clearly seen that even for denser atmospheres of the order of several bars, the $\Delta T$ value required for agreement in precipitation between the atmospheric model and the network data (i.e., precipitation $\geqslant$1.5\,mm d$^{-1}$, see Fig. \ref{comp}) is still of the order of 40-50\,K. How could such high temperatures be achieved?

\subsubsection{Numerical uncertainties}

Firstly, atmospheric models, hence predictions of surface temperature and precipitation, are subject to (numerical) uncertainties.

An important aspect is the radiative transfer, hence the strength of the greenhouse effect. Usually, atmospheric models do not solve the radiative transfer equation with a line-by-line code to save computational speed. They instead use faster approximations such as exponential sums or correlated-k methods (e.g., \citealp{wiscombe1977}, \citealp{goodyyung1989}). It can be shown that such methods are quite accurate (e.g., \citealp{goody1989}, \citealp{west1990}, \citealp{lacis1991}), and most atmospheric models compare well to high-resolution radiative transfer calculations (e.g., \citealp{goldblatt2009}, \citealp{wordsworth2010cont}, \citealp{ramirez2014}). The impact on surface temperature is probably small, of the order of 1-2\,K at most. The opacity databases which are typically employed, such as Hitran and Hitemp, have evolved quickly in recent years (e.g., \citealp{rothman2009}, \citealp{rothman2013}), and the effect on radiative-convective calculations can be quite important (e.g., \citealp{pavlov2000},  \citealp{kratz2008}, \citealp{kopparapu2013}). Especially the treatment of line and continuum absorption of CO$_2$ is another uncertainty factor for early Mars calculations (e.g., \citealp{halevy2009}, \citealp{wordsworth2010cont}, \citealp{mischna2012}). Numerical issues, such as the vertical discretization of the atmospheric column, can also introduce uncertainties in surface temperature of a few K for optically thick atmospheres ($p_{CO2} \geq$0.2-0.5\,bar). In the model used here, a warming effect of up to 7\,K was observed upon increasing the number of levels in the troposphere where the opacity is largest. This effect is comparable to other 1D radiative-convective models (e.g., \citealp{tian2010}, \citealp{kopparapu2013}). For atmospheres with less CO$_2$ partial pressure, the vertical discretization is not important for the calculation of surface temperatures. In summary, we should probably assign an inherent uncertainty of 5-10\,K to the calculation of mean surface temperatures. Figure \ref{deltat_contour} suggests that by using these uncertainty estimates, the discrepancy between geological data and atmospheric model estimates could increase or decrease by about 20\,\%.

\subsubsection{1D vs 3D models}

The 1D atmospheric model used in this work calculates annual mean, global mean temperatures. However, the networks mostly formed at lower latitudes where conditions were likely warmer than such global mean values. Indeed, 3D modeling studies (e.g., \citealp{wordsworth2013}, \citealp{urata2013}) have suggested that the latitudinal temperature gradient in mean surface temperature is of the order of up to 20-30\,K. However, this is below the needed threshold inferred from Fig. \ref{deltat_contour}. 

Seasonal or obliquity effects could be much stronger (see e.g. Fig. 3 in \citealp{wordsworth2013}), with temperature effects of the order of 50-60\,K in some locations. Therefore, a viable alternative to a continously wet early Mars are scenarios where obliquity cycles allow for periodic warming at the networks' geographic location which would then allow for formation and precipitation of liquid water \citep{wordsworth2013}. This issue should be investigated with geological constraints on the valley network formation timescales. These are poorly constrained, but it seems that at least a somewhat sustained presence of water was required to form them \citep{ansan2013}.

We acknowledge the importance of full 3D atmospheric studies, demonstrated for many Solar System objects, e.g. for Earth, Mars, Titan etc. However, 1D and 3D studies can be used complementarily to investigate different aspects of atmospheric processes. For example, 1D models of Titan have been used to tackle specific problems of the thermal structure (e.g., \citealp{mckay1989titan}) or haze formation (e.g., \citealp{lavvas2008}, \citealp{lavvas2008titan_profiles}), whereas 3D models of Titan are used to address especially the problem of (equatorial) superrotation (e.g., , \citealp{hourdin1995}, \citealp{friedson2009}, \citealp{newman2011}, \citealp{lebonnois2012}). For early Mars, 1D-2D models have been used to investigate mean surface temperatures and assess the validity of greenhouse solutions (e.g.,  \citealp{postawko1986}, \citealp{poll1987}, \citealp{kasting1991}, \citealp{mischna2000}, \citealp{Cola2003}, \citealp{tian2010}, \citealp{fairen2012}, \citealp{vparis2013marsn2}). 3D models have demonstrated the importance of obliquity and transport effects (e.g., \citealp{johnson2008}, \citealp{forget2013}, \citealp{wordsworth2013}, \citealp{mischna2013}, \citealp{urata2013}, \citealp{scanlon2013}).

It is clear that 3D models are physically more consistent than 1D models since they better resolve the planetary surface, include clouds and horizontal as well as vertical energy transport by winds or tracer species such as water. However, they introduce many parameters, some of them even on a sub-grid scale, hence parameterizations are needed (e.g., leaf area index, \citealp{fraedrich1999}, or surface roughness, \citealp{wordsworth2013}). Other parameters include, e.g., precipitation efficiencies or precipitation thresholds, cloud properties and the amount of cloud nuclei (e.g., \citealp{wordsworth2013}). Most of them are not known for early Mars.
 
On applying 3D models, one should always bear in mind that the boundary conditions might so uncertain that one could be simulating conditions which are very different.

In contrast, 1D models somewhat keep the number of parameters under control. In our model as used for the current manuscript, effectively, the only parameters are the choice of surface albedo and the choice of the RH profile.

In addition, the current spatial latitude-longitude resolution of most 3D GCM studies of early Mars (e.g., 32x24 or 32x32 in \citealp{wordsworth2013} and \citealp{forget2013}) is insufficient to resolve individual networks. Therefore, even 3D GCM simulations would probably not be enough to address the question of local precipitation for individual networks.

As stated in the Introduction, we aim at assessing the general strength of the early Mars hydrological cycle, and not individual network channels. Therefore, the choice of a 1D model is not thought to severely impact our conclusions.

Ideally, for Early Mars, one would apply both 3D snapshots of the obliquity cycle to investigate the detailed feedback mechanisms, as well as 1D studies to investigate a wide parameter range (of e.g. atmospheric composition) not possible in 3D. In our present work, a 1D study is used to provide only basic indications of the hydrologic cycle, which in turn depend on obliquity etc. Such a two-pronged approach (i.e. complimentary 1D and 3D studies) could in our opinion prove beneficial for Early Mars.

\subsubsection{Atmospheric composition}

In the global mean, \citet{vparis2013marsn2} suggested that 500\,mbar of N$_2$ would provide about 10\,K surface warming. Other candidates for providing surface warming are additional greenhouse gases such as CH$_4$ and SO$_2$ (e.g., \citealp{postawko1986}, \citealp{yung1997}, \citealp{johnson2008}, \citealp{tian2010}, \citealp{mischna2013}). However, an unresolved question is the possible formation of aerosols due to high SO$_2$ concentrations. Like O$_3$ (a possible by-product of CO$_2$ photochemistry, e.g. \citealp{selsis2002}), SO$_2$ is a strong UV absorber. This could help to avoid the surface cooling due to CO$_2$ Rayleigh scattering observed at high surface pressures, since SO$_2$ or O$_3$ absorption bands would strongly reduce the planetary albedo (e.g., \citealp{vparis2013ice}). Furthermore, together with near-IR CH$_4$ absorption, such trace gases would probably warm the lower stratosphere such that CO$_2$ cloud formation would be largely inhibited, thus reducing the cloud radiative forcing. Volcanic gases such as H$_2$ could also play a role in warming early Mars since collision-induced absorption could provide a large greenhouse effect, and Rayleigh scattering of H$_2$ is far less efficient than for CO$_2$ (e.g., \citealp{stevenson1999}, \citealp{pierrehumbert2011hydro_hz}, \citealp{wordsworth2011}, \citealp{wordsworth2013earth}, \citealp{ramirez2014}). The study of \citet{ramirez2014} found mean surface temperatures at high CO$_2$ partial pressures which are compatible with the network data presented here.

\subsection{Uncertainties in the estimation of discharge and runoff}

Our estimates of discharge and the inferred precipitation rates are subject to large uncertainties. The simple assumption that discharge rates can be used to directly estimate precipitation rates neglects many important factors that control this relationship. In this section we address some of the uncertainties that are necessarily involved in attempts to reconstruct the water budget in a 3.8 Gyr-old catchment.

\subsubsection{Discharge calculations}

Previous studies used different approaches to estimate paleodischarges of Martian channels. Ideally, reliable estimates of paleodischarge would require not only the measurement of morphometric properties of channels, but also the knowledge of or at least well-constrained assumptions of other, independent parameters such as particle size distributions of mobilized sediment and channel floor roughness (see review by \citealp{kleinhans2005}). The input parameters for such micro-scale or hydraulic methods are usually poorly known for modern Mars, since they cannot be easily obtained from orbit. However, they can be measured in situ by rovers such as MSL/Curiosity, which landed on the distal parts of an alluvial fan in Gale Crater. For example, the size distribution of rounded and fluvially transported particles was determined by \citet{yingst2013}. The flow velocity, water depth, and discharge of channels that transported the particles that now form outcrops of conglomerate was estimated by \citet{Williams2013science}. Nevertheless, the data limitations for most other Martian channels imply that macro-scale or hydrologic methods need to be applied for paleodischarge estimates (see \citealp{burr2010}). Such form-discharge approaches are based on morphological parameters that are relatively simple to measure and have been used extensively (e.g., \citealp{irwin2005}, \citealp{williams2013icarus}). It is important to note, however, that gravity effects (e.g., on channel geometry) and the unknown strength of channel banks on Mars introduce significant errors to our paleodischarge estimates \citep{williams2013icarus}. Moreover, as mentioned, since we usually cannot measure the channel width, we use an empirical relation of channel width to valley width (see Eq. \ref{valleyw}).

From eq. \ref{runoff}, the uncertainty $\frac{\Delta R}{R}$ of the calculated runoff is given by

\begin{equation}
\label{deltar}
\left(\frac{\Delta R}{R}\right)^2=\left(\frac{\Delta C}{C}\right)^2+\left(\frac{\Delta Q}{Q}\right)^2\end{equation}

The uncertainty $\frac{\Delta Q}{Q}$ is partly due to the measurement error $\frac{\Delta W_V}{W_V}$ associated with the channel or valley width. However, a very significant part is also contributed by uncertainties due to the application of the approximative eqs. \ref{qmean} ($U_6$, for deriving discharges) and \ref{valleyw} ($U_{10}$, for deriving channel widths). Combining these, we find

\begin{equation}
\label{deltaq}
\left(\frac{\Delta R}{R}\right)^2=\left(\frac{\Delta C}{C}\right)^2+\left(1.71\cdot \frac{\Delta W_C}{W_C}\right)^2+U_6^2+U_{10}^2
\end{equation}

For a conservative estimate of a 20\,\% error in the measurements of both catchment area and channel width, and taking $U_6$=0.79 (standard error for applying the power law of eq. \ref{qmean} to discharge calculations, Table 3 in \citealp{osterkamp1982}) and $U_{10}$=0.1 (interquartile range for the proportionality constant in eq. \ref{valleyw}, \citealp{penido2013}), we thus assign a 1\,$\sigma$ error of 90\,\% to the calculated runoff. Note that this neglects any potential systematic uncertainty from applying eq. \ref{valleyw}, implying that our runoff uncertainty estimates are probably optimistic.

\begin{figure}[h]
  \centering
  \includegraphics[width=250pt]{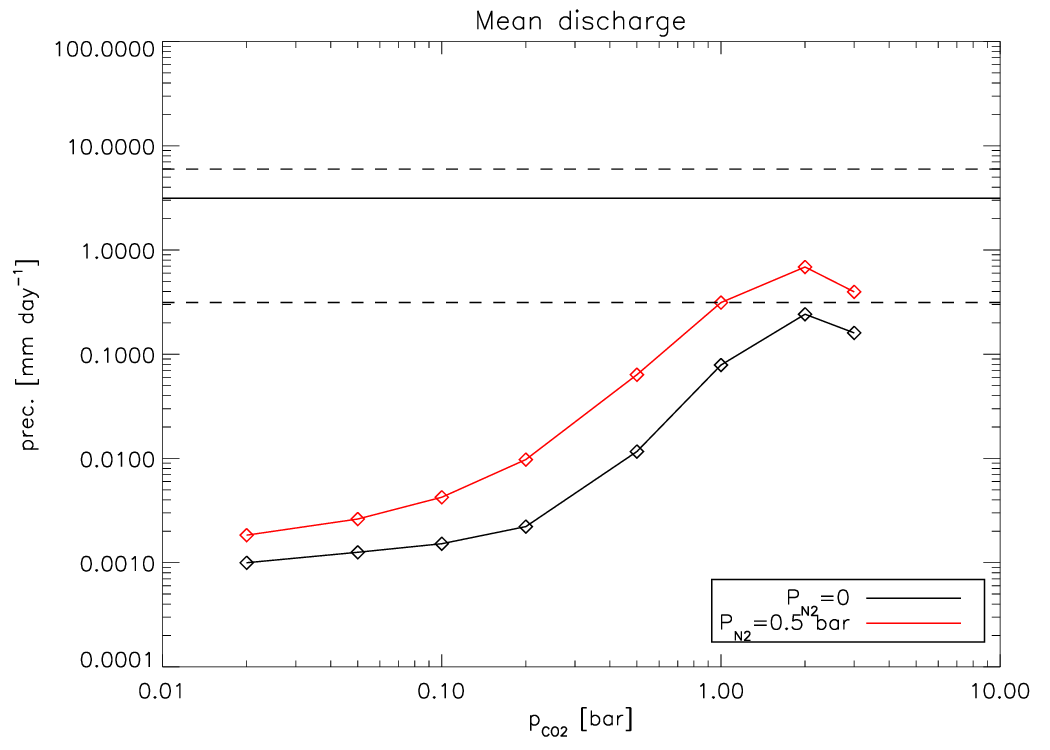}\\
  \caption{Median network runoff (plain line) with its 1\,$\sigma$ error of 90\%, (dashed lines) compared to total model precipitation (\,mm d$^{-1}$) with nominal values of RH=1 and $pr$=13.3\%. Black: No N$_2$, red: 0.5\,bar N$_2$.}\label{ratioerr}
\end{figure}

Figure \ref{ratioerr} shows the effect of this error estimate on the comparison between atmospheric model precipitation and derived network runoff. For high-pressure atmospheres with significant amounts of N$_2$, both data sets seem to compare roughly within error bars. Note, however, that our 1D steady-state model atmospheres predict snowfall rather than rainfall (see surface temperatures in Fig. \ref{ts}), hence, the associated runoff would be zero.

Another error source stems from the fact that the approximation formulae provided by \citet{osterkamp1982} were derived for the Missouri river basin. For other basins, such formulae will probably different, although, as stated by \citet{osterkamp1982}, a large number of individual river channel were included in the analysis. Therefore, data are expected to cover a wide range of channel widths and discharge values, and the total scatter is probably not much larger than the already stated 79\%.

\subsubsection{Role of evaporation and infiltration}

The derivation of precipitation rates from channel discharge rates (assuming the channel banks are full) is far from being straightforward. First, the area of the catchment must be known, the shape of which may have been changed by erosion or tectonic deformation in the last $\sim$3.8 Gyr. Additional uncertainties may arise from data limitations during the mapping process \citep{penido2013}. Second, drainage in a channel does not directly correspond to runoff or precipitation in its catchment area. If discharge is used to infer precipitation rates, the relative roles of evaporation and infiltration need to be considered (transpiration does not exist on Mars since there is no vegetation). In general, the water balance in a catchment system can be expressed as

\begin{equation}\label{balance}
  \Delta S = P-Q_S-Q_G-ET
\end{equation}

where $\Delta S$ is the change in the amount of water stored in the system, $P$ is rainfall, i.e. liquid water precipitation, $Q_S$ is surface discharge, $Q_G$ is groundwater discharge, and $ET$ is evapotranspiration. For practical purposes, the amount of water in a system is often assumed to be constant. In terrestrial hydrological studies, infiltration is often considered to be negligible \citep{anderson2010geo}, since the infiltrated water returns to the stream system as groundwater. If the overall climate is cold \citep{wordsworth2013} and the substrate in which a channel is incised is subject to permafrost, the infiltration rate would even tend towards zero, because a permafrost ground would, to a first order, be impermeable (see also \citealp{heldmann2005}). Applying the same logic to Mars, we can thus consider Eq. \ref{balance} to be equivalent to

\begin{equation}\label{marsbalance}
  \int (R-E)dt=\frac{1}{A}\int Q_S dt
\end{equation}

with $R$ is the rainfall rate, $Q_S$ the hydrograph and $A$ the area of catchment. Evaporation and sublimation rates on modern Mars have been estimated by several past studies. \citet{chittenden2008} determined sublimation rates of pure water ice as a function of temperature, wind speed, and relative humidity. They found that temperature is the main factor influencing the sublimation rate. The results of \citet{chittenden2008} show sublimation rates of less than a millimeter per hour (see Table 2 of \citealp{chittenden2008}). Whereas these results apply to water ice, \citet{sears2005} performed laboratory experiments with a 7\,mbar CO$_2$ atmosphere at 0\,$^{\circ}$C and obtained evaporation rates of 1.01$\pm$0.19\,mm h$^{-1}$. After a correction for gravity effects, \citet{sears2005} predict evaporation rates on Mars of 0.73$\pm$0.14\,mm h$^{-1}$. If one assumes that runoff took place in a rather cold environment on early Mars \citep{wordsworth2013}, then both sublimation of ice and evaporation of liquid water may have occurred. The effect of thin water ice layers with temperatures around the freezing point of water were investigated by \citet{moore2006}. These authors find evaporation rates of 0.84$\pm$0.08\,mm h$^{-1}$ and 1.24$\pm$0.12\,mm h$^{-1}$ with and without a thin ice layer, respectively, and conclude that the presence of thin water ice layers does not have a significant effect on the evaporation rates. The corresponding values for a whole day are 2.02 and 2.98\,cm d$^{-1}$ with and without a thin ice layer, respectively. These rates are of the same order of magnitude as those obtained from discharge estimates, and therefore have to be taken into account and need to be added to the discharge-derived rate before calculating precipitation rates. Of course, the catchment would not be a free surface of water, in which case evaporation rates may be somewhat lower than those obtained for free water surfaces, but in any case the amount of evaporated water would not be negligible.

Another question is whether the assumption of insignificant infiltration in Eq. \ref{balance} is valid. The upper crust in the southern highlands, where the valley networks are observed, probably consists of heavily cratered regolith. Such substrate would be characterized by high infiltration capacities that may have impeded runoff production (e.g., \citealp{baker1986}, \citealp{gulick1990}, \citealp{grant1993}, \citealp{carr2000}, \citealp{irwin2008} and references therein). In a study of a terrestrial analogue on Hawaii at the Kilauea and Ka\`{u} deserts, which are characterized by brecciated basaltic material, \citet{craddock2012} estimate that runoff can only be generated if precipitation rates are $>$2.5\,cm h$^{-1}$. Infiltrated water would feed groundwater aquifers, which in turn may have contributed to the generation of sapping valleys by groundwater seepage. Groundwater flow, however, may have been different to surface runoff and its associated catchment areas, and it is unclear how to quantify this effect. Since the mean annual temperatures indicated by our study (Fig. \ref{ts}) and previous modeling of early Mars temperatures \citep{wordsworth2013} are below 0$^{\circ}$C, we assume permafrost conditions in our hydrologic estimates and therefore neglect the effects of infiltration, including the possible (but minor) infiltration of snow melt into frozen ground. 

The above comparison between geomorphologic data and atmospheric models (see e.g., Fig.\ref{comp}) explicitly assumes no infiltration and no evaporation, i.e. all precipitated water would feed the valley network runoff. Even under such optimistic assumptions, there is a disagreement between both methods. Hence, it is clear that a more detailed analysis, taking into account the full eq. \ref{balance}, would most likely lead to a larger disagreement. However, given the uncertainties in runoff estimates (see Fig. \ref{ratioerr}), a full hydrological analysis is probably not warranted by the current data quality.

\subsection{Dry warm Mars?}

\label{ocean}

Mean surface temperatures above freezing (warm Mars) do not necessarily lead to high precipitation (wet Mars). Paleoclimate studies of the Earth during phases of large super-continents imply that the continental climate during these phases might have been very dry \citep{parrish1993paleo}. This suggests that another condition for extensive precipitation on early Mars would be the presence of large standing bodies of liquid water. 3D climate studies of, e.g., \citet{soto2012} suggest that without a large ocean, early Mars would have been extremely arid even if it was warm. The existence of a Noachian ocean is currently debated. Several studies however advocate that the northern lowlands have been at least partly ocean-covered (e.g., \citealp{clifford2001}, \citealp{perron2007}, \citealp{dibiase2013}, \citealp{deblasio2014}). If indeed such a large ocean would have existed, precipitation on a warm early Mars would have been abundant \citep{soto2012}.

\subsection{Limits to evaporation}

\label{limit_evap}

Liquid-water precipitation must be balanced by evaporation of surface water into the atmosphere, otherwise the water column will deplete quickly. Evaporation then leads to a cooling of the surface through the release of latent heat  (e.g., \citealp{fraedrich1999}). In the model used here, this latent heat flux $F_{\rm{lat}}$ is neglected, and the surface energy balance is only determined by the radiative flux $F_{\rm{rad}}$. However, for increasing amounts of evaporation, the contribution of $F_{\rm{lat}}$ can become significant, and the surface energy balance should read as

\begin{equation}
\label{surfacebalance}
F_{\rm{surf}}=F_{\rm{rad}}+F_{\rm{lat}}
\end{equation}

The radiative flux in eq. \ref{surfacebalance} is determined by the net stellar flux reaching the surface and the radiative longwave cooling ( i.e., downwards-upwards longwave fluxes). It is calculated directly in the atmospheric model.

%\begin{equation}
%\label{fluxess}
%F_{\rm{rad}}=F_{\ast,\rm{net}}+F_{\rm{IR,net}}=(F^{\rm{dn}}_{\ast}-F^{\rm{up}}_{\ast})+(F^{\rm{dn}}_{IR}-F^{\rm{up}}_{IR})
%\end{equation}

 %\textbf{The stellar upward flux $F^{\rm{up}}$ determined by the planetary surface albedo ($F^{\rm{up}}=A_S\cdot F^{\rm{dn}}$, $A_S$ set constant to 0.21, see \citealp{vparis2013marsn2}). The thermal upward flux $F^{\rm{up}}_{IR}$ is calculated following the Stefan-Boltzmann law. For high-pressure atmospheres, the greenhouse effect becomes strong, and $F_{\rm{IR,net}}$ is close to 0 (see \citealp{vparis2013marsn2}). For surface pressures higher than about 0.5\,bar, the surface radiative flux is almost entirely dominated by the stellar component in eq. \ref{fluxess}.}

For high-pressure atmospheres, the greenhouse effect becomes strong, and the longwave cooling is close to 0 (see \citealp{vparis2013marsn2}). For surface pressures higher than about 0.5\,bar, the surface radiative flux is almost entirely dominated by the stellar flux.

\begin{figure}[h]
  \centering
  \includegraphics[width=150pt]{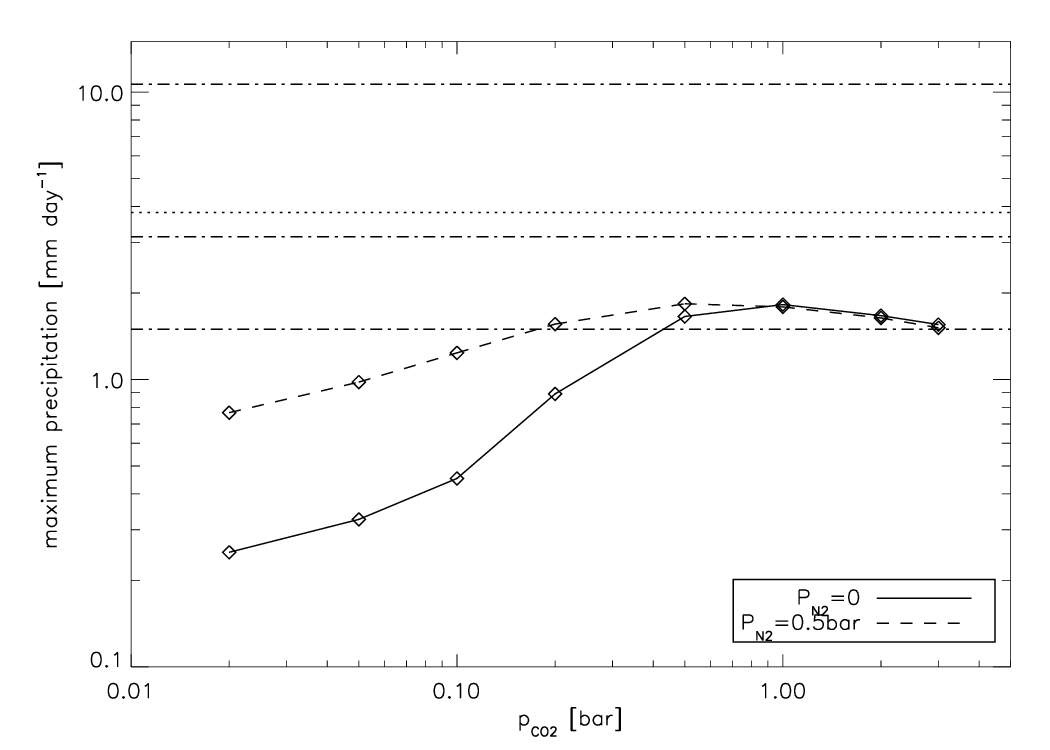}
    \includegraphics[width=150pt]{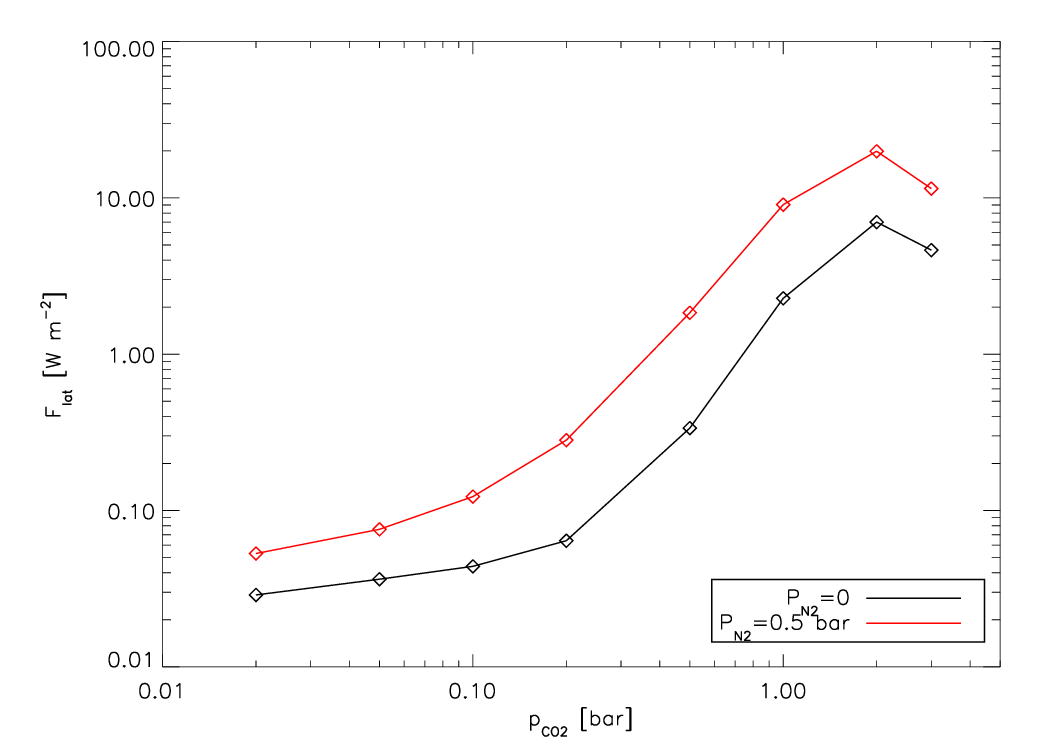}\\
  \caption{Left: Maximum global mean precipitation based on surface energy balance. Network maximum, minimum and median indicated by dot-dashed horizontal lines. Dotted horizontal line indicates maximum precipitation for zero-albedo atmosphere. Right: Latent heat flux of precipitation rates from Fig. \ref{comp}.}\label{maxprec}
\end{figure}

The right panel of Fig. \ref{maxprec} shows the latent heat flux of the precipitation rates as taken from Fig. \ref{comp}, assuming a latent heat of 2,500\,J\,g$^{-1}$. It is clearly seen that except the high-CO$_2$ runs ($p>1$\,bar), the value of $F_{\rm{lat}}$ is somewhat negligible, justifying the model assumption of taking only the radiative flux into account.

In an approach similar to \citet{ogorman2008}, the left panel of Fig. \ref{maxprec} shows the maximum of global mean precipitation, if the entire net surface radiative flux of the model atmospheres would be used to evaporate water. Also indicated are the values of precipitation from Table \ref{tabledata}. Furthermore, the dotted horizontal line shows the absolute global mean maximum of precipitation ($\approx$3.8\,mm\,day$^{-1}$), if the entire top-of-atmosphere incoming stellar flux ($\approx$109\,W\,m$^{-2}$) would be available for surface water evaporation. This is equivalent to assuming a zero-albedo atmosphere (no absorption or scattering of radiation), which is clearly not realistic, as indicated in Fig. \ref{ts}. Furthermore, longwave cooling as well as surface reflectivity would have to be zero as well, which would be mutually inconsistent. In addition, such a scenario would need a global ocean which is questionable (see discussion above in Sect. \ref{ocean}).

In summary, these calculations suggest that the maximum global mean precipitation sustainable under the faint-Sun conditions assumed for early Mars is about 1.8\,mm\,day$^{-1}$ (Fig. \ref{maxprec}, left panel). This is still much less than inferred precipitation from the valley network analysis. Accounting for a latitudinal or seasonal gradient in precipitation compared to a global mean (see discussions above and eq. \ref{enhance}), this could have been enough to sustain valley network formation, even if barely for the highest-runoff channels. However, given that such an amount of precipitation requires that all of the stellar energy reaching the surface be converted into latent heat, these highest-runoff channels were probably not formed during long, continously wet periods.

\section{Conclusions}

\label{concl}

In this work, we have estimated runoff rates for 18 Mars valley networks and compared them to estimated precipitation rates from an atmospheric model of early Mars. Model atmospheres were composed of varying amounts of CO$_2$ and N$_2$. We also varied parameters such as atmospheric relative humidity and precipitation efficiency to estimate their impact on precipitation values.

Runoff rates inferred from valley network data and precipitation from the atmospheric model generally disagree by about an order of magnitude for high-pressure CO$_2$ atmospheres, the runoff rates being much larger than the atmospheric model precipitation rates. At low CO$_2$ partial pressures, a scenario favored by escape models, the discrepancy is even larger. This suggests that early Mars was probably not a continously wet environment. Rather, these results point to sporadic periods of high precipitation, probably due to a change in obliquity, or, more likely, (local) melting events where a large snow/ice reservoir accumulated over a geologically long time (e.g., an obliquity cycle) could have melted in a short period. Even though such a conclusion is not new, it is the first time that both geological and atmospheric modeling approaches have been used to test quantitatively  the hypothesis of wet versus dry early Mars.

Geological runoff rates are uncertain by currently at least 90\,\% and atmospheric models are also subject to uncertainties. Therefore, quantitatively comparing the estimates of the amount of precipitated water from both approaches is challenging. However, since global mean surface temperatures as calculated by our atmospheric modeling are far below freezing, calculated model precipitation is most likely not a good predictor of valley network runoff. This points the direction of future research, i.e. the need to better constrain estimates of paleo-discharges in ancient river networks on early Mars, determine the water source for runoff (e.g., snowmelt or rain) and its timescales and refine mesoscale atmospheric modeling of early Mars climate before more rigorous conclusions can be drawn.

Nevertheless our study represents an important first step in constraining precipitation on early Mars involving both the atmospheric and geological communities.

\section*{Acknowledgements}

This study has received financial support from the French State in the frame of the "Investments for the future" Programme IdEx Bordeaux, reference ANR-10-IDEX-03-02. This research has been supported by the Helmholtz Association through the research alliance "Planetary Evolution and Life". We thank two anonymous reviewers for stimulating comments. Discussions with Mareike Godolt, Franck Selsis and Franck Hersant are gratefully acknowledged. Discussions with Robert Craddock have sharpened our view on infiltration rates on Mars.

\bibliographystyle{natbib}
\bibliography{atmos_geo_arxiv}

\end{document}